\newif\ifcomments

\commentsfalse 

\newif\ifcorrections
\correctionstrue

\documentclass{aa}  
\usepackage{graphicx}
\usepackage{txfonts}
\usepackage{color}
\usepackage{natbib}
\usepackage{subcaption}
\usepackage{gensymb}

\newcommand{\Msol}{\mbox{ M}_\odot}

\newcommand{\Gs}{\mbox{ Gs}}

\newcommand{\yr}{\mbox{ yr}}
\newcommand{\kyr}{\mbox{ kyr}}
\newcommand{\pc}{\mbox{ pc}}
\newcommand{\au}{\mbox{ au}}

\newcommand{\rsink}{r_\textrm{sink}}

\begin{document} 

\title{Jets and outflows of massive protostars}

\subtitle{From cloud collapse to jet launching and cloud dispersal}

\author{        A. K\"olligan
                \inst{1}
                \and
        R. Kuiper\inst{1}
        }

\institute{Institute of Astronomy and Astrophysics, University of T\"ubingen,
                Auf der Morgenstelle 10, D-72076 T\"ubingen\\
                \email{anders.koelligan@uni-tuebingen.de}
        }

\date{Received 20/06/2018; accepted 24/10/2018}

\abstract{
        Massive stars live short but intense lives. While less numerous than low-mass stars, they enormously impact their surroundings by several feedback mechanisms. They form in opaque and far-away regions of the galaxy, such that one of these feedback mechanisms also becomes a record of their evolution: their bright large-scale jets and outflows.
}
{
In a comprehensive convergence study, we investigate the computational conditions necessary to resolve (pseudo-) disk formation and jet-launching processes, and analyze possible caveats. We explore the magneto-hydrodynamic (MHD) processes of the collapse of massive prestellar cores in detail, including an analysis of the forces involved and their temporal evolution for up to two free-fall times.
}
{
        We conduct MHD simulations using the state-of-the-art code PLUTO, combining nonideal MHD, self-gravity, and very high resolutions as they have never been achieved before. Our setup includes a $100\Msol$ cloud core that collapses under its own self-gravity to self-consistently form a dense disk structure and launch tightly collimated magneto-centrifugal jets and wide-angle tower flows.
}
{
        We show a comprehensive evolutionary picture of the collapse of a massive prestellar core with a detailed analysis of the physical processes involved and our high-resolution simulations can resolve a magneto-centrifugal jet and a magnetic pressure-driven outflow, separately. 
        The nature of the outflows depends critically on spatial resolution. Only high-resolution simulations are able to differentiate a magneto-centrifugally launched, highly collimated jet from a slow wide-angle magnetic-pressure-driven tower flow. Of these two outflow components, the tower flow dominates angular-momentum transport. The mass outflow rate is dominated by the entrained material from the interaction of the jet with the stellar environment and only part of the ejected medium is directly launched from the accretion disk.
A tower flow can only develop to its full extent when much of the original envelope has already dispersed.
        Taking into account both the mass launched from the surface of the disk and the entrained material from the envelope, we find an ejection-to-accretion efficiency of $10\%$.
        Nonideal MHD is required to form centrifugally supported accretion disks and the disk size is strongly dependent on spatial resolution. 
        A converged result for disk and both outflow components requires a spatial resolution of $\Delta x \leq 0.17 \au$ at $1\au$ and sink-cell sizes $\leq 3.1\au$.
}
{
Massive stars not only possess slow wide-angle tower flows, but also produce magneto-centrifugal jets, just as their low-mass counterparts.
The actual difference between low-mass and high-mass star formation lies in the `embeddedness' of the high-mass star which implies that the jet and tower flow interact with the infalling large-scale stellar environment, potentially resulting in entrainment. }


\maketitle

\section{Introduction}

Many details of the formation process of massive stars are still poorly understood. This is mainly due to their large average distances from our solar system and due to the fact that their early evolutionary stages take place in opaque clouds of gas and dust barely penetrable by our current observational capabilities \citep[e.g.,][]{Zinnecker2007}. Nevertheless, we can observe a prominent feature of their evolutionary phase: their bright large-scale outflows \citep[e.g.,][]{Frank2014}.

While massive stars are rare, they have a much more significant impact on their surroundings than the large number of low-mass stars. During their whole lifetime, their feedback mechanisms influence not only their close neighborhood but their whole natal clusters. Therefore, they influence the initial mass function (IMF) of their home clusters, provide heavy elements for later generations of stars, and act as giant stellar laboratories (such as G023.01-00.41 \citep{Sanna2014}).

The enormous radiative and mechanical luminosities of massive stars impact a vast range of scales and processes, such as the reionization of the universe, the evolution of galaxies, the regulation of the interstellar medium, the formation of star clusters, and even the formation of planets around stars in such clusters.

The knowledge gained through the study of low-mass stars possibly provides important insights into massive star formation as well, although which aspects of their evolution appear similarly in massive stars it is still
a matter of debate.
Some aspects of star formation should theoretically work similarly, irrespective of scale. These processes are called scale invariant. 

Although, we study outflows within a collapsing cloud core here, that is, including the nonscale-invariant self-gravity of the gas, the gravitational potential of the launching and collimation region is dominated by the central star, in which case the magneto-hydrodynamic (MHD) equations are to a major degree invariant with respect to the gas density.
The aspects of MHD that govern the launching and collimation of outflows are self-similar. 
Even in this case, however, the environmental conditions can be very different and result in different initial conditions, as in the case of low- and high-mass star formation.

To form massive stars, an equally massive mass reservoir is needed, which is then dominated by gravity and, depending on the initial mass-to-flux ratios\footnote{This is usually given in units of the critical mass-to-flux ratio (Eq. \eqref{eqn:CritMass2Flux}). A mass-to-flux ratio of $1$, in principle, means that the Lorentz force can balance gravity in the initial cloud and so the core is sub-critical and will not collapse.}, the magnetic field. Thermal pressure, on the other hand, is (in contrast to low-mass star formation) expected to be negligible in comparison with gravity, centrifugal, and magnetic forces, especially related to outflow driving. On the other hand, thermal pressure is expected to be the primary vertical support of the accretion disk and is consequently required for a realistic disk scale height.

For massive cores, feedback by radiation pressure, ionization, and line-driving potentially contribute immensely to the evolution, while for low-mass cores, these effects are not present or relevant at all.
While in both cases the early evolution takes place in deeply embedded regions, in low-mass cores, the envelope is to a large degree depleted before their radiation becomes a relevant factor for their continuing evolution. This is not the case for massive cores where the strong gravitational pull quickly accumulates mass, and with it potential energy that is converted into thermal and radiative energy, accompanied by high accretion luminosities. When the protostar finally starts its fusion processes, its feedback mechanisms eventually stop accretion in the nonshielded polar regions completely, and launch winds and so on. All these processes influence the evolution of the collapse immensely while a significant part of the envelope is still present and the (proto)star is still not observationally accessible.

Outflows potentially have different launching principles and work on different spatial and temporal scales. There are numerous known ways to launch outflows: they can be accelerated by line driving \citep{Kee2016, Vaidya2011}, by magneto-centrifugal acceleration at the interface of the stellar magnetosphere with the disk \citep[X-Winds][]{Shu1994}, by magneto-centrifugal processes from the centrifugally supported part of the disk \citep{Blandford1982}, by magnetic pressure alone \citep{Lynden-Bell2003}, by continuum-radiation pressure \citep[e.g.,][]{Yorke2002,Kuiper2010,Kuiper2011}, and on the largest scales by expanding HII regions \citep[e.g.,][]{Kuiper2018a}.
A comprehensive overview of the state-of-the-art of modeling MHD jets in massive star formation is given in Section \ref{MHDsims}.

In this paper, we focus on the very early evolutionary stages of massive star formation and are interested mainly in MHD launching processes without contributions from radiation. This means that we focus on magneto-centrifugal launching from rotating accretion disks and winds driven by magnetic pressure. In the early stages that we simulate, processes like line-driving, radiation pressure, and ionization are considered to be negligible, though in principle they may become important during the temporal span of our simulations.
There are established analytic models describing these types of outflows: the fast jet that is launched from the rapidly rotating disk inside of the Alfv\'en-radius\footnote{The Alfv\'en-radius is the radius at which magneto-acoustic waves propagate as fast as the material flows in the polar direction. It is an elaborate way to differentiate between the regions where the material flow governs the magnetic field topology and the region where the magnetic field topology governs the flow direction. We note the keyword \textit{polar direction}; this means here that, while in the toroidal direction the magnetic field lines are dragged along with the flow, in the polar direction the rotating magnetic field lines move material around and can, therefore, also accelerate material. 
} described by \citet{Blandford1982} and a slower, more massive wind launched outside the Alfv\'en-surface where the magnetic field is wound up by differential rotation, producing a so-called magnetic tower flow as described by \citet{Lynden-Bell2003}.

In the following, we use the term {outflow} as a general term for the more specific terms {jet} and {wind}. We use {jet} to refer to a  fast ($>10$ km/s), magneto-centrifugally driven outflow launched from the very central few astronomical units of the accretion disk. 
        {Wind} on the other hand is used to refer to a slow ($<10$ km/s) magnetic-pressure-gradient-driven outflow. 
        To differentiate both types of outflows in the simulational data, we use two primary criteria: the velocity and the ratio of the toroidal to the poloidal magnetic field strength, with the jet-launching region having a higher polar magnetic strength, accelerating the material to high velocities and the wind launching region having a high toroidal magnetic field strength and the accelerated gas having a much lower terminal velocity. A suited criterion would also be the position of the launching area of the outflow with respect to the Alfv\'en surface, as the magnetic pressure gradient needed for the magnetic-pressure-driven wind can only develop if the inertial forces of the differentially rotating gas can wind up the magnetic field.

Beside the physical launching process itself, its influence on the evolution of the system composed of star, accretion disk, and the envelope is important for our understanding of star formation. Outflows could alleviate another important issue: as \citet{Mestel1965} first realized, prestellar clouds have a much higher average angular momentum than the final stars. Therefore, there must be an efficient way to remove excess angular momentum during the evolutionary process. Magnetic breaking and magnetically induced turbulence due to the magneto-rotational instability can potentially provide this mechanism \citep{Shu1994,Lovelace1995}. Magnetically driven winds can remove a large portion of angular momentum without removing too much mass due to the strong coupling between particles in the jet and the rest of the accretion disk through magnetic fields. This also means that jets and outflows inject (angular) momentum into their environment and are an important feedback effect that influences not only a single star but a whole star-forming region.

In section \ref{sec:previous_studies}, we summarize important observations of massive protostars and the state-of-the-art of MHD modeling of jets and outflows from massive protostars (including the cloud collapse). In section \ref{sec:ac_methods}, we describe the methods and parameters used in our simulation. In section \ref{Temporal_Evolution}, we then give an overview on the temporal evolution of our collapse simulations. In sections \ref{Convergence_res} and \ref{Convergence_sinks}, we analyze the convergence properties of our numerical setup with respect to the resolution and the sink-cell size and discuss caveats. In section \ref{Physical_Effects}, we describe our physical findings in detail, explaining disk formation and outflow mechanisms, in section \ref{caveats}, we discuss influences of higher magnetic field strength as a possible caveat of the physical parameters of our setup, and in section \ref{sec:summary} we summarize our results.

\section{Previous studies}
\label{sec:previous_studies}

In the following, we give a brief overview of previous studies on jets and outflows in massive star formation. We start with observed properties and continue with a comprehensive review of computational studies that include magnetic fields. Therein, we highlight results pertaining to the disk formation process, the launching of outflows, and their characteristic properties, such as disk mass, radius, and outflow momenta.

\subsection{Observed properties of jets and wide-angle outflows}

A very comprehensive review is given in \citet{Frank2014}. Here, we simply want to mention a few important results that can be easily compared with simulations.

While direct observation of the outflow launching engine is not yet possible, there are several other quantities that can be derived from larger-scale observations and are, therefore, accessible to us.
One of those quantities is the mass outflow rate for which all necessary components can be derived from other directly measurable quantities. This means that typical outflow rates from massive prestellar cores are well known, as well as the accretion-to-outflow efficiency.
Classical T-Tauri stars \citep{Cabrit2007,Agra-Amboage2009} as well as Herbig Ae/Be stars \citep{Ellerbroek2013} have accretion-to-outflow rates of about $10\%$.
It is not entirely clear if this ratio also holds for more massive O/B stars but if these massive stars have similar accretion and ejection mechanisms they are expected to have similar rates. This ratio is also predicted in the seminal work by \citet{Blandford1982} on self-similar magneto-centrifugal jets, though they actually predict a dependence on the magnetic lever arm and one arrives at this $10\%$ ratio only for the very idealized conditions that they assume in their model.

Directly related to the accretion-to-outflow-efficiency is the issue that the observed star formation efficiency is only on the order of $\simeq 0.3$ \citep{Offner2014, Padoan2014} which means that there must be a way to remove mass from the collapsing cloud core or halt it for extend timescales (longer than the evolutionary timescale of the forming star).
A possible resolution to this issue is, as several authors have pointed out, that the observed accretion rates are far too low to acquire the masses observed in
the IMF \citep{Evans2009, CarattioGaratti2012} which
means that accretion (and outflows) are thought to be episodic events. 
Recent simulations and observations \citep{CarattioGaratti2016, Stecklum2017b} of accretion disks in massive star formation provide ample evidence that these episodes are triggered by the accretion of gaseous clumps created by disk fragmentation \citep{Meyer2017}, similar to FU Orionis objects in lower-mass star formation.

A very remarkable recent observation by \citet{McLeod2018} indicates a massive young stellar object in the Large Magellanic Cloud (LMC) that shows signs of such a highly collimated fast magneto-centrifugal jet as describe here. In its surroundings, nearby O-stars seem to have dispersed the envelope by alleviated ionization feedback enabled by the low dust content of the LMC, revealing for the first time the inner workings of a protostellar outflow of a massive protostar. They report jet velocities of $300-400$ km/s, mass outflow rates of $2.9\times 10^{-6}\Msol$ yr$^{-1}$, and derive outflow lifetimes of $(28-37)\kyr$ from the extent of the jet. In the following section, we summarize the results of relevant numerical studies.

\subsection{Magneto-hydrodynamic simulations in high-mass star formation}
\label{MHDsims}

There have been several studies on outflows of low-mass stars \citep{Tomisaka2002, Hennebelle2008, Machida2008, Price2012, Machida2013, Tomida2013, Machida2013, Bate2014, Tomida2015}, though simulations of massive stars and especially collapse simulations of massive prestellar cores are much rarer.
In terms of massive cloud-collapse simulations, we want to mention an early simulation by \citet{Banerjee2007} who start their simulation with a massive Bonnor-Ebert-sphere setup and compare three simulations: an isothermal collapse, a collapse with radiative cooling, and a magnetized collapse with radiative cooling. Their study is relevant here because they demonstrated that only a simulation including a magnetic field could self-consistently produce outflows. They also argue, though they do not simulate this, that these outflows could help to relieve the radiation-pressure-problem of massive star formation by channeling radiation out of their outflow cavities. Also, given the computational resources of that time, they were only able to resolve a magnetic tower flow, and did not see signs of a magneto-centrifugal jet.

Generally, convergence considerations are rare, or at least they are not mentioned. There are three studies on collapse simulations that compare results for different resolutions: \citet{Hennebelle2011} and \citet{Seifried2011, Seifried2012}.
\citet{Hennebelle2011} simulated a massive ($100\Msol$) cloud collapse in ideal MHD. They ran their adaptive mesh refinement (AMR)-simulations with a maximal resolution of $2$ au and compared these to a run with a lower resolution, corresponding to a maximal refinement of $8$ au. They focus their attention and discussion on the influences of the magnetic field on cloud evolution and find episodic slow-velocity outflows.

The series of simulations conducted in \citet{Seifried2011} and \citet{Seifried2012} have similar initial conditions to our own simulations but have a different computational focus. They use a three-dimensional (3D) grid and do not include nonideal MHD effects. Both papers describe the collapse of $100\Msol$ cores with variable rotational and magnetic energy densities using AMR with a maximal resolution of $4.7\au$.
Thematically, \citet{Seifried2011} focus on accretion rates and centrifugally supported disk formation, finding that strong magnetic fields can suppress centrifugally supported disk formation for simulations with mass-to-flux ratios of less than $10$. \citet{Seifried2012} analyze in detail the conditions for magneto-centrifugal jet launching by \citet{Blandford1982} and develop and test a criterion to decide whether the jet is launched centrifugally or magnetically. They find that both processes are present, and that the central part of the outflow is primarily centrifugally launched and the outer outflow is launched to a larger degree by magnetic forces.
Moreover, they find that with high magnetic fields the collimation of their outflows is rather poor due to the strongly sub-Keplerian disk rotation that does not build up a strong toroidal magnetic field necessary to collimate the jet.

\citet{Machida2014} give a comprehensive overview comparing different collapse simulations with respect to sink-cell size, resolution, and initial conditions for low and massive prestellar clouds. They tried to reproduce the simulations by several other authors, with the single study involving massive stars included in their reproduction being the one by \citet{Seifried2011,Seifried2012}. Interestingly,  they repeated the  simulations of \cite{Seifried2011} in a nested grid of dimensions $64\times64\times32$ with equatorial symmetry, resulting in higher resolutions of up to $0.6$ au.
The most remarkable finding of their study is that the outflow properties are greatly affected by sink treatment and the size of the sink cell in all of their simulations.

Other authors focus on the issue of fragmentation without analyzing outflows in detail. Fragmentation in this context means that the Jeans mass can be reached at several small independent regions. This, in turn, leads to the fragmentation of a single massive prestellar core into several small gravitationally bound objects, effectively suppressing the formation of a single more massive entity.
The interaction of these fragments then also hinders the formation of stable and fast outflows, as shown in \citet{Commercon2011} and \citet{Hennebelle2011}.
Both groups conduct MHD-AMR-collapse-simulations of $100\Msol$ cores with a maximal resolution of $2$ au.
\citet{Hennebelle2011} focus on the effects of magnetic fields in massive star formation and fragmentation for different mass-to-flux ratios ($\bar \mu= 120, 5, 2$) with a barotropic equation of state. Due to fragmentation, they were only able to find slow episodic nonbipolar outflows with average velocities  $\simeq 3-4$ km/s and maximal velocities of $40$ km/s.
In contrast, \citet{Commercon2011} focus on the suppression of fragmentation due to interplay of magnetic and radiative forces. They mention that in a simulation with $\bar \mu=2$, they form only a single fragment with an outflow velocity of $2$ km/s.

Several other authors are mostly interested in the large-scale environment and the effects of feedback of massive star formation. The simulations by \citet{Wang2010} and \citet{Myers2014a} simulate the collapse of massive clouds with RMHD but use observationally motivated subgrid-models to introduce outflows from their sink cells and do not self-consistently produce outflows. They use large-scale computational domains focusing on the influences of these outflows on the large-scale magnetic field structure, star cluster formation, and star formation rate, thereby deducing the CMF and IMF of whole clusters.
\citet{Peters2011} are also focussed on the larger scales and simulate a $1000\Msol$ with AMR-RMHD and compare simulations with and without magnetic fields to study the effect of magnetic fields. 
They are able to produce very-low-velocity outflows self-consistently. They interpret their results as implying that large-scale low-velocity tower flows of massive cores are much weaker than expected from their low-mass counterparts and they reason that this is mainly due to fragmentation; though to our knowledge, this could also be an effect of their large, $98$ au cell sizes (as stated above, they focus on larger scales). Therefore, their simulations do not resolve the launching region of the centrifugally launched fast jet component and even the lower-velocity tower flows could reach higher velocities if they were launched closer to the central mass.

A recent study that specifically discussed jets and outflows of massive protostars is \citet{Matsushita2017}. They show 3D nested grid simulations of massive prestellar cores with a fixed resolution but varying energy ratios using a barotropic equation of state. They follow the outflow for a few $10^4$yr and focus on the mass ejection rate.
They do include nonideal MHD effects and use a nested grid approach with a fixed number of cells in each level of $64\times64\times32$ combined with equatorial symmetry. On the highest level, they achieve a cell size of just $0.8$ au which can resolve the launching region quite well, though is nearly one order of magnitude lower than the resolution we reach in the launching region. Similar to our simulation, they use a sink cell with a radius of $1\au$ . The main finding is that a proportionality between accretion and outflow rates exists in their simulations and they conclude that low- and high-mass stars probably have a similar launching mechanism, which means that their outflows are launched by magneto-centrifugal and magnetic-pressure-driven mechanisms.
While their simulations are relatively similar to ours, they do not discuss how their outflows are launched and accelerated. In particular, no distinction between a magneto-centrifugal and a magnetic tower flow is made in the interpretation of their results, although they state that the outflow was driven by a magneto-centrifugal process.
They focus on the relation of the resulting properties, like outflow momentum, accretion rate, and so on, to physical input parameters such as initial cloud mass, magnetic field, and angular velocity, among others, while we concentrate on the physical launching mechanisms and how numerical parameters influence the evolution of the cloud core.
Finally, they do not consider the effects of different resolutions or sink-cell sizes. Studies that do consider resolution aspects are discussed in the following section.

\subsection{Convergence aspects}

As already mentioned, only a few authors test their setups on convergence and even if they do, only \citet{Seifried2011} compare more than two resolutions. Therefore, it remains unclear if many of those simulations converge to a stable solution and which resolution is necessary.

\citet{Seifried2011} repeated their simulations at resolutions  a factor of four higher as well as lower than their fiducial case. They compare the density and temperature profiles as well as the velocity in the midplane. In the higher-resolution runs they can resolve shock features in density and temperature to a better degree than in their low-resolution runs and they state that the vertical structure of the disk may not even be sufficiently resolved in their highest resolution. Also, their criterion for sink-particle creation is met more often for higher resolutions.

Additionally, they show the accretion rate of their sink particle for the first $2\kyr$ of their simulations. The simulation with the highest resolution clearly shows a significantly different behavior compared to the lower-resolution simulations.
During the last $500\yr$ shown, a process seems to evolve that is only resolved by the highest resolution. 
Also, there is no indication of a trend of convergence. During these latter $500\yr$, the accretion rate in the highest-resolution simulation lies in between the low- and the mid-resolution simulations and, on average, increases with time. Judging from this trend, we expect the average value of the accretion rates to diverge significantly for longer simulations and with it the mass in the sink.

This underlines the strong dependence of the physical evolution of the system under investigation on numerical resolution.
With their convergence study focussed on the disk structure, it is not clear if they meet the requirement to resolve the jet-launching region enough for a quantitative analysis of the outflow properties.
We also want to point out that they introduce a threshold for the minimal density of $1.78\times 10^{-12}\textrm{ g cm}^{-3}$ in their simulations to limit the maximal velocity of magneto-acoustic waves in their computational domain, which in turn results in larger numerical time steps.
This, together with their relatively low resolution, could act as an effective magnetic dissipation which would then lead to an underestimation for the minimal mass-to-flux ratio to build up centrifugally supported disks.

While \citet{Hennebelle2011} study convergence properties of their simulations by comparing a high- and a low-resolution run, they use two slightly different Riemann solvers with different numerical diffusivities and they run the high-resolution simulation only for half the time they run the low-resolution simulation ($0.1-0.2$ free-fall times after the formation of the first protostar for the high-resolution simulations, and $0.4-0.5$ free-fall times for the low-resolution run). They find significant differences between the two simulations and remark on caveats for the low-resolution runs.

In comparison, our simulations have very high resolutions with a minimal cell size of just $0.09\au$ and small sink cell radii of only $1.0\au$. Simulations with this resolution have never been done before and enable us to analyze the details of disk formation and outflow launching in detail.
After we give an overview on the physical processes of our collapse simulations in Sect. \ref{Temporal_Evolution}, our detailed convergence studies in Sects. \ref{Convergence_res} and \ref{Convergence_sinks} show that many physical processes in such simulations are extremely dependent on  resolution and sink-cell size.
To this end, we compare the evolution of disk properties (disk radius, disk mass, and disk lifetime), outflow properties (linear and angular momenta, ejection-to-outflow ratios), and protostellar mass and accretion rate for simulations with different resolutions ($\Delta x = 0.09$, $\Delta x = 0.17$, $\Delta x = 0.37$) and different sink-cell sizes ($\rsink=1.0\au$, $\rsink=3.1\au$, $\rsink=10.9\au$, $\rsink=30.3\au$).
Finally, in Sect. \ref{Physical_Effects}, we analyze the physical structure of the disk and outflow in detail, using the forces and magnetic field topology to produce a comprehensive physical picture of the processes shown. First, though, we introduce our numerical methods.

\section{Methods}
\label{sec:ac_methods}

The basis of our simulations is the modular MHD code PLUTO \citep{Mignone2007} combined with a self-gravity solver developed by \citet{Kuiper2010}.
PLUTO is a free and open-source code for the solution of mixed hyperbolic/parabolic systems of partial differential equations, primarily intended for (but not limited to) the use in astrophysical fluid dynamics using a finite volume or finite difference approach based on Godunov-type schemes. It natively includes modules for MHD and relativistic MHD.

We make use of its HLLD Riemann solver with linear interpolation and integrate by Runge-Kutta method of order 2.
Since numerical schemes do not preserve the solenoidality of the magnetic field naturally, PLUTO includes several methods to ensure it. Here, we utilize PLUTO's state-of-the-art constrained transport algorithm to this end.

To close the system of equations, we chose an isothermal equation of state.
PLUTO's modular nature allowed us to implement equations of state with different adiabatic indices that could also be varied depending on the position, time, and local variables to mimic the effects of radiative heating/cooling in a controlled way. While this approach might be useful in the future, it basically pre-determines a significant part of the simulation. Therefore, we conclude that an ideal equation of state combined with radiative transfer would be the \textit{ultima ratio} with respect to a realistic outcome of our collapse simulation, as well as the only way to derive a realistic evolution for longer time spans. This would, on the other hand, vastly increase the number of physical processes and significantly aggravate the analysis of the MHD effects we want to focus on.
Therefore, an isothermal equation of state represents a solid choice for analyzing MHD effects in early phases of the collapse of massive prestellar cores and we will include radiative transfer in future studies.

PLUTO also supports the treatment of dissipative effects like viscosity, magnetic resistivity, and thermal conduction. In our simulations, we only need the former two.
For the shear viscosity of the disk, we adopt the prescription by \citet{Kuiper2010}.

We use axisymmetric and  mirrorsymmetric boundaries at the polar axis and the equator, respectively, as well as zero-gradient boundary conditions at the inner radial boundary.
The outer radial boundary uses zero-gradient boundaries as well but with the radial velocity constrained to positive, that is, outflowing, velocities.

Our simulations include ohmic resistivity as a phenomenological model for the magnetic dissipation in the dead, dense, and radiatively shielded region of the disk, thereby enabling the formation of a centrifugally supported accretion disk that is not destroyed by magnetic breaking.
We implement the resistivity model used by \cite{Machida2007a} which is based on a numerical study by \cite{Nakano2002}. These latter authors investigated the different mechanisms of magnetic flux loss in molecular clouds by drift of dust grains. The resulting ohmic resistivity of 
\citet{Machida2007a} is given by
\begin{equation}
        \centering
        \eta = \frac{740}{X_\mathrm{e}}\sqrt{\frac{T}{10\mathrm{K}}}\left[1-\tanh{\frac{n}{10^{15}\mathrm{cm}^{-3}}}\right]\mathrm{cm}^2\mathrm{s}^{-1}
        \,,
  \label{eqn:machida2007-model}
\end{equation}
with the ionization degree
\begin{equation}
        \centering
        X_\mathrm{e}=5.7\times 10^{-4} \left(\frac{n}{\mathrm{cm}^{-3}}\right)^{-1}
        \,.
  \label{eqn:ionization}
\end{equation}
We use the same equation at an isothermal temperature of $10$ K.
The dense dead zone in the accretion disk of the  cores of the massive clouds that we simulate regularly reaches densities of $n_\mathrm{H} > 10^{12}$ cm$^{-3}$. This is according to \cite{Nakano2002}, the regime where ohmic dissipation is the dominant flux-loss mechanism. Three aspects of our numerical setup are discussed in more detail: the grid in Sect.. \ref{sec:Grid}, the initial conditions in Sect. \ref{sec:init_cond}, and the Alf\'en velocity limiter in Sect. \ref{sec:AlfvenLimiter}.

\subsection{Grid}
\label{sec:Grid}

We run our simulations on a 2D spherical grid with equatorial- and axal symmetry. By stretching the grid in the radial direction logarithmically and tuning the number of cells in radial and polar direction accordingly, we generate a grid with cells of equal spatial extent in the radial and the polar direction that naturally reduces the cell size close to the origin. This increases the resolution of our simulations in the central region, where eventually the outflows launch and where the physical processes in the disk and outflow cavity require the most resolution. This simple but effective grid geometry has the highest dynamic range among all regular grid techniques used in previous studies of MHD jets in massive star formation. To achieve a similar dynamic range in a Cartesian AMR simulation with 126 cells (corresponding to the highest number of cells we use in radial direction), one would need more than 13 levels of AMR refinement (in addition to the base resolution) and with it great computational resources.
In our simulations, the entire computational grid reaches from the central sink cell radius, which is a parameter that we vary in our convergence study (correspondingly, $r_\mathrm{sink} = 0.1\au,0.3\au,1.0\au,3.1\au,10.9\au,30.3\au$), up to $0.1\pc$. 

\subsection{Initial conditions}
\label{sec:init_cond}

Initially, the cloud core has a radially symmetric density distribution $\rho \propto r^{-1.5}$ with the rotational axis and grid axes aligned along with the direction of the uniform magnetic field. Kinetic and thermal energies are chosen such that the core is supercritical, that is, its gravitational pull is stronger than the forces stabilizing it and collapses immediately.
The core rotates initially with solid body rotation of a frequency of $3\times10^{-13} $s$^{-1}$, corresponding to a total kinetic to gravitational energy ratio of $2\%$. The thermal energy is, with a (isothermal) temperature of just $10$ K, negligible (thermal to gravitational energy fraction $=0.5\%$).

Considering only gravity, the core would collapse with a free-fall time of
\begin{equation}
  t_\mathrm{ff} = \frac{\pi}{2}\frac{r^{3/2}}{\sqrt{2GM}} \simeq 52\kyr
        \label{eqn:fft}
        \,.
\end{equation}
The mass-to-flux ratio $\bar \mu$ in units of the critical mass-to-flux-ratio \citep{Nakano1988} is
\begin{equation}
        \bar \mu = \left. \frac{M_\mathrm{core}}{\Phi_\mathrm{core}} \middle / \left( \frac{M}{\Phi} \right)_\mathrm{crit} \right.
        = \left. \frac{M_\mathrm{core}}{\int B_\mathrm{z} dA} \middle/ \frac{1}{2\pi\sqrt{G}} \right.\,,
  \label{eqn:CritMass2Flux}
\end{equation}
with $M_\mathrm{core}$ being  the total core mass, $\Phi_\mathrm{core}$  the total magnetic flux, and the area $A$  set to $20$.

\subsection{Alfv\'en limiter}
\label{sec:AlfvenLimiter}

The maximal time step of hydrodynamic simulations following explicit Godunov-type schemes like ours is, depending on the local cell size, limited by the highest signal velocity anywhere on the grid. In our case, this velocity is either the maximal velocity of material streams or magneto-acoustic waves as well as ordinary sound waves. The velocity of magneto-acoustic waves is limited by the sound speed or the Alfv\'en velocity (whichever is higher).

The Alfv\'en velocity can get particularly high in environments with high magnetic field strengths and low densities. In collapse simulations like ours, a situation like this occurs naturally in the jet-launching region. This is because material falling in from above can be accreted onto the star traveling along the magnetic field close to the pole without any resistance by magnetic or centrifugal forces. The magnetic field, on the other hand, is constantly accreted through the infalling material in the equatorial plane. 
Therefore, the magnetic field piles up in the center while the bulk mass above the disk is accreted by the protostar. This is even amplified when an outflow is launched above the star, as the outflow cavity reduces the density even further.
These evolutionary processes lead to an increase of magnetic field and decrease of density. Therefore, the Alfv\'en velocity will increase quickly. As a result, the time step of our simulation decreases to what is effectively a halt.

Therefore, we chose to increase the density in this region to limit the maximal Alfv\'en velocity to $700$ km/s. We tested various velocity limits and $700$ km/s constitutes a trade-off between realistic magnetic field evolution and reasonable time steps. We monitor the mass that we create in this way throughout the whole simulation, and in all simulations it remains below $6\%$ of the initial mass until one free-fall time.
We continue with an overview of the various processes that occur during the collapse of a massive prestellar core in the following section.

\section{Temporal evolution}
\label{Temporal_Evolution}

In the following, we describe the most important evolutionary phases of the system in chronological order. The timescales refer to the time from the onset of cloud collapse to the simulation with a sink-cell size of $1\au$ and with a resolution of $126\times20$ grid cells.
Qualitatively similar results also appear for lower-resolution simulations and larger sink-cell sizes, only shifted in time. Differences between these runs are discussed in Sects. \ref{Convergence_res} and \ref{Convergence_sinks}, respectively.

\begin{figure}[h]
  \centering
  \includegraphics[width=0.5\textwidth]{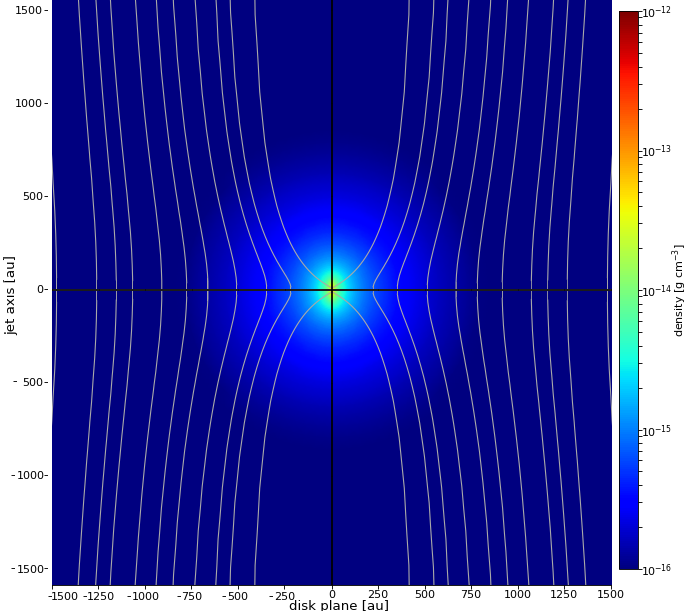}
        \caption{Density distribution near the protostar after $3.5\kyr$ ($0.5\kyr$ previous to jet launching), overlaid with the hourglass-shaped magnetic field lines.}
  \label{fig:hour-glass}
\end{figure}

In the first $4 \kyr$, the gravitationally supercritical cloud core collapses under its own gravity and neither magnetic nor centrifugal forces reach comparable magnitudes.
The infalling material flows with super-Alfv\'enic velocities down to the equatorial plane. The comparatively low magnetic field strength has nearly no influence on the flow in this epoch and the magnetic field topology is determined primarily by the gas flow. As a result, the magnetic field gets dragged into the center along with the gas, yielding the well-known hourglass-like shape (Fig. \ref{fig:hour-glass}).
While most of the material of the original envelope is still infalling from all directions, angular-momentum conservation leads to the formation of a notably rotationally flattened envelope and hereupon a disk-like structure that is not yet centrifugally or magnetically supported. 

The magneto-centrifugal jet launches at $\simeq4\kyr$. It immediately clears out an outflow cavity with a density that is  two orders of magnitude lower than the neighboring infalling envelope.
Directly after the initial jet launching, magnetic and centrifugal forces can withstand gravity in the midplane for the first time and the accretion disk grows continuously outwards.
\begin{figure}
\centering
\includegraphics[width=0.49\textwidth]{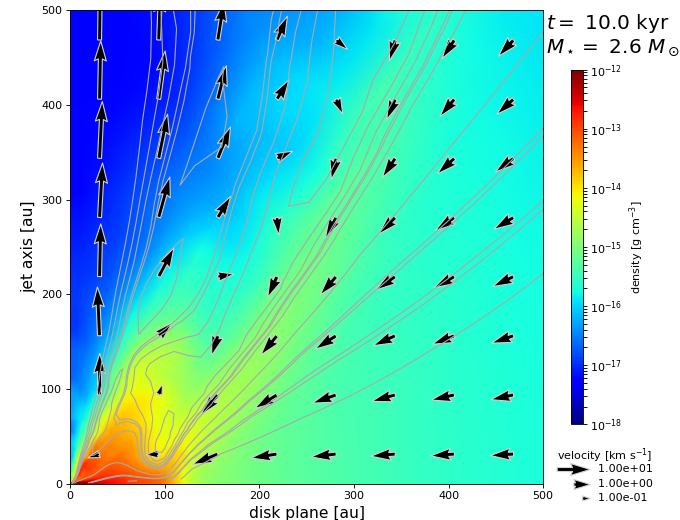}
\caption{Increased density distribution above the disk after $10\kyr$, overlaid with magnetic field lines. 
}
  \label{fig:accretion_clump}
\end{figure}
The increasing magnetic field strength and the strong ram pressure of the outflow hinder accretion within the polar outflow cavity. At the same time, there is still a flow of material falling downwards along the cavity wall. This congestion leads to an increased density above the disk that is threaded and supported by the magnetic field (Fig. \ref{fig:accretion_clump}).

At $\simeq 8\kyr$, a separate launching/acceleration region becomes distinguishable from the previously present magneto-centrifugal jet.

This acceleration takes place at larger radii and higher altitudes above the disk (compared to the fast initial jet) in a region of relatively high density (compared to the outflow cavity) and high toroidal magnetic field strength. Therefore, it can be identified as a magnetic tower flow.
The tower flow's acceleration in this direction is not as strong as the acceleration of the magneto-centrifugal component and results in lower flow velocities. The magneto-centrifugal jet reaches outflow velocities of several $100$ km/s, while typical outflow velocities of the tower flow only reach a few km/s.

From this point in time on, the accretion disk grows in size, the density contrast between the disk and its dense atmosphere above the disk continuously increases and the outflow is continuously active. The launching region of the fast magneto-centrifugal jet expands towards higher altitudes and the launching region of the slower tower flows towards higher altitudes and to larger disk radii (See Sect. \ref{Physical_Effects} for details).

When the simulation reaches one free-fall time, the mass reservoir above the protostar gets depleted and with it, the supply of angular momentum for the disk ceases as well. Then the only channel left for accretion is through the accretion disk. As a result, the fast magneto-centrifugal jet component vanishes.
When infall ceases, both the reservoir of material for star formation and its associated ram pressure are removed.
Without this pressure from above, the magnetic pressure gradient becomes stronger than gravity up to much larger radii, and, as a result, the launching region of the slow tower flow vastly expands; within the next $20\kyr$, it encompasses $10000\au$ and continues to grow to the full extent of our simulated region of $0.1\pc$ before the end of the simulation at $100\kyr$. 
As all processes described here are potentially dependent on the details of our numerical setup, our convergence considerations are presented in the following two sections.

\section{Convergence properties for different resolutions}
\label{Convergence_res}

In this section, we present and discuss the convergence properties of the simulations performed.
Specifically, we check here in detail if the results of the simulation change with resolution and/or the size of the central sink cell. Such numerical convergence tests are crucial to extract robust scientific results from numerical simulations in general. Only parameters and results which show a proper convergence behavior represent meaningful outcomes of such a study.
This section is subdivided into the physical objects that are analyzed, since different physical effects potentially have different resolution requirements.
First, we discuss the influences of different resolutions for the evolution of the accretion disk. Subsequently, we continue with the outflows as well as the protostar. In the sections following that, a similar analysis is carried out for different sink-cell sizes.

For the resolution analysis, we performed three simulations with $126\times20$ cells, $64\times10$ cells, and $32\times5$ cells in the radial, correspondingly polar direction. 
All physical and numerical parameters were kept exactly the same for the three simulations. Most importantly, their grid always stretches from $1.0\au$ to $20626.5\au = 0.1\pc$.
The given cell numbers translate to the following resolutions $\Delta x$ in the innermost regions close to the protostar and where the jet is launched:
\begin{table}[h]
\centering
\label{tab:res_sims}
\begin{tabular}{llllll}
        cell number & $\Delta x$ [au] \\
        \hline
$126\times20$ & 0.09  \\
$64\times10$  & 0.17  \\
$32\times5$   & 0.37  \\
\end{tabular}
\caption{Cell numbers and corresponding resolutions $\Delta x$ used in the simulations.}
\end{table}

\subsection{The accretion disk}

\begin{figure*}[h]
\centering
\begin{subfigure}[b]{0.49\textwidth}
        \includegraphics[width=\textwidth]{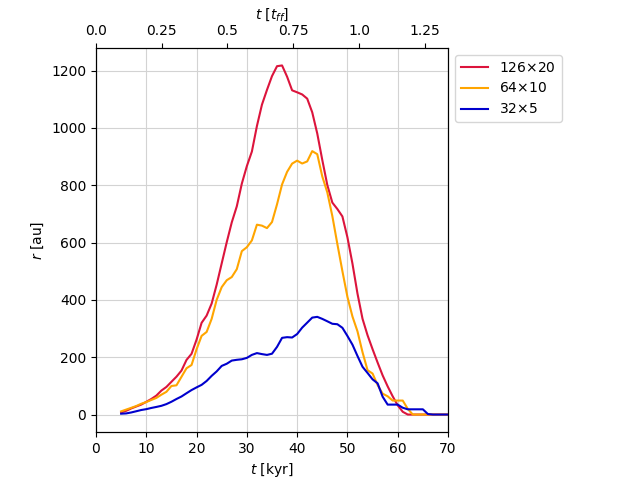}\\
        \caption{} 
  \label{fig:conv_diska}
\end{subfigure}
\begin{subfigure}[b]{0.49\textwidth}
        \includegraphics[width=\textwidth]{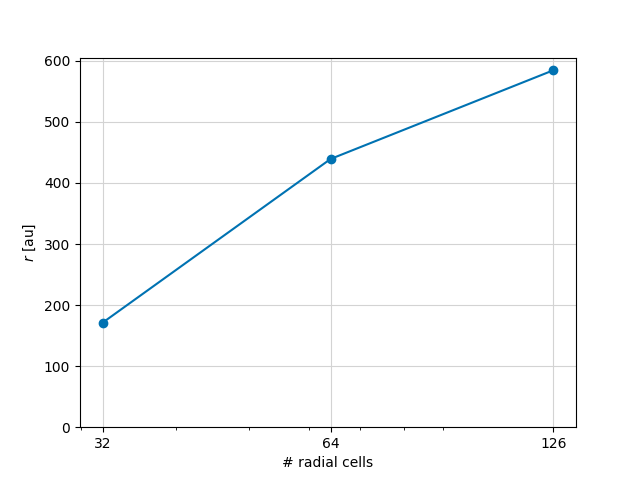}\\
        \caption{} 
  \label{fig:conv_diskb}
\end{subfigure}
\caption{Convergence properties of the disk radius for different resolutions. (a) Disk radius as a function of time. (b) Time-averaged disk radius as a function of resolution.
}
\label{fig:conv_disk}
\end{figure*}

Generally, we find circumstellar disk formation in all simulations performed, regardless of resolution or sink-cell size. Critical for this outcome is nonideal MHD which reduces magnetic breaking in the magnetically decoupled, dense inner part of the accretion disk. Also, we found very massive and spatially extended disks that reach radii of several hundreds of astronomical units, which are, to our knowledge, the most extended disks ever reported in MHD simulations.

The physics of the accretion disk is converged for all but the lowest-resolution case, in the sense that we do not observe any principle evolutionary feature present in one but not the other simulation. This is also reflected in the plots shown in Fig. \ref{fig:conv_disk}; in panel (a), we present the evolution of the disk radius as function of time and, in panel (b), we show its average value over the course of the first free-fall time. We also examined the disk mass as a function of time and, consequently, as a criterion for its convergence. However, it evolves in a strikingly similar manner to the radius of the disk and is, therefore, not shown here.
To plot the disk radius, we flag a region as part of the disk when it reaches a density threshold of $10^{-15}$ g cm$^{-3}$ and if it is in a relatively stable force equilibrium. To check the latter, we compare the centrifugal force in the region with the gravitational pull and accept it as being in equilibrium if the centrifugal force lies within $\pm5\%$ of the gravitational force.
We double-checked this criterion in all simulations and it reliably only detects parts of the stellar environment that can be clearly identified as part of the accretion disk, that is,  regions lying close to the equatorial plane and with significant density contrast to the envelope.

The most striking similarity between different kinds of simulations is the (centrifugally supported)  lifetime of the disk. The first detectable centrifugally supported structure lies within a $100\yr$-interval for all resolutions, and after $\simeq 50\kyr$, they vanish similarly within a time span of $\simeq 5\kyr$.

It is quite obvious that the disk in the lowest-resolution simulation evolves differently from that in the higher-resolution simulations, with a significantly smaller and less massive disk, also showing an earlier peak radius (and mass).
This is also reflected in the disk's time evolutions in panel (b), with much larger discrepancies between simulations $32\times5$ and $64\times10$ than between $64\times10$ and $126\times20$. 
The disk radius increases very significantly from the simulation $32\times5$ to $64\times10$, with a difference of $61\%$, while from $64\times10$ to $126\times20$, the difference reduces to just $34\%$. This trend is indicative of more closely approaching a converged result.
Also, the time evolution of the higher-resolution simulations clearly shows a pronounced correlation during the whole simulation.

Additionally, a possible influence on the disk radius in the highest-resolution simulation is due to it having more cells in the polar direction, resulting in the ability to resolve the same scale height with more cells. Therefore, the density of the cells at the equatorial plane tends to increase with resolution and with that, there tend to be more cells that fulfill the density criterion that we apply at each cell (in the equatorial plane, where we measure the disk radius and where most of the mass is concentrated). As a result, we expect the actual physical properties of the disk to be globally better converged than shown by these plots.

Concluding, both high-resolution simulations with $64\times10$ and $128\times20$ grid cells yield converged results with respect to the accretion disk physics. This grid relates into spatial resolutions of $0.17\au$ and  $0.8\au$, respectively, within the launching region close to the central host star.

\subsection{The magneto-centrifugal jet and the magnetic tower flow}
\label{sec:conv_outflows_res}

\begin{figure*}[h!]
\centering
\begin{subfigure}[b]{0.49\textwidth}
        \includegraphics[width=\textwidth]{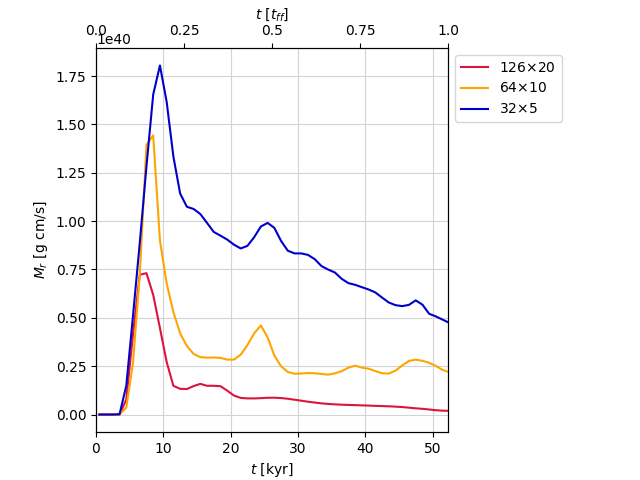}\\
        \caption{} 
\end{subfigure}
\begin{subfigure}[b]{0.49\textwidth}
        \includegraphics[width=\textwidth]{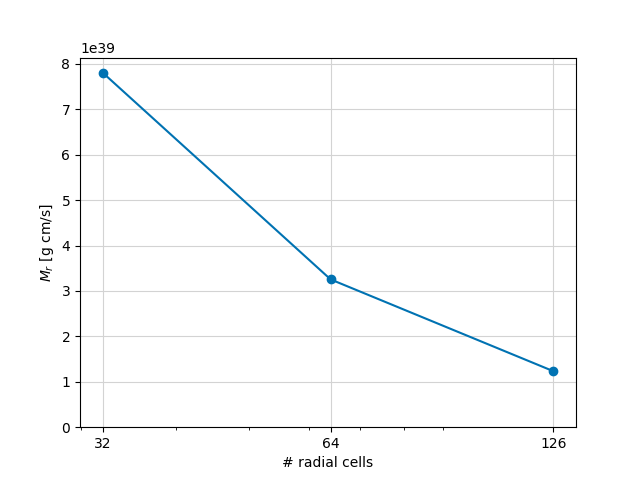}\\
        \caption{} 
\end{subfigure}
\begin{subfigure}[b]{0.49\textwidth}
        \includegraphics[width=\textwidth]{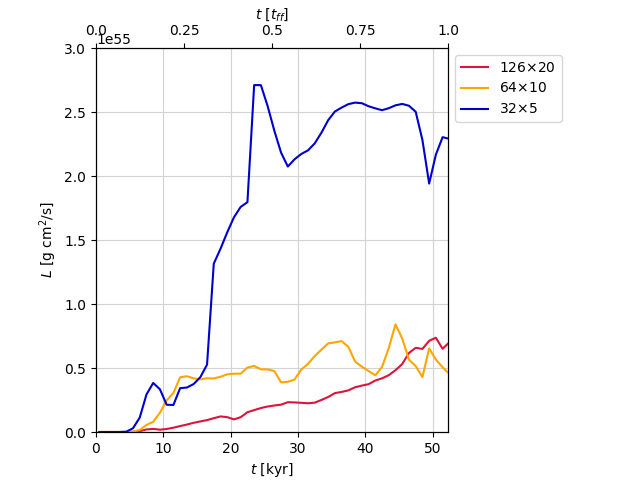}\\
        \caption{} 
\end{subfigure}
\begin{subfigure}[b]{0.49\textwidth}
        \includegraphics[width=\textwidth]{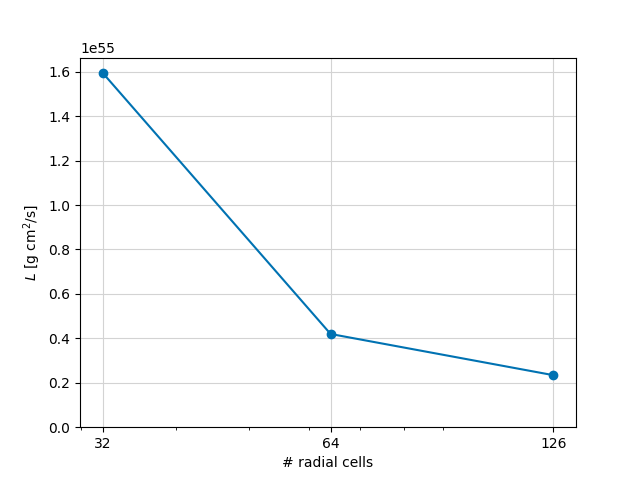}\\
        \caption{} 
\end{subfigure}
\caption{Convergence properties of the outflow for different resolutions.
	(a) Total linear outflow momentum as a function of time.
	(b) Time-averaged outflow momentum as function of resolution.
	(c) Total outflow angular momentum as a function of time.
	(d) Time-averaged outflow angular momentum as function of resolution.
}
\label{fig:conv_outflow}
\end{figure*}

In this section, we analyze the convergence properties of the outflows. Comparable to the previous considerations on the disk, we apply a criterion to decide whether a region is considered part of the outflow or not. Here, we allow only regions within $30\degree$ of the polar axis, a minimum vertical velocity of $0.1$ km/s, and with a distance of $>300\au$ from the origin to contribute to the total mass, momentum, or angular momentum of the outflow. The minimal radius and maximal polar angle are used to only include stable parts of the outflow, thereby excluding short-lived perturbations in the disk or the central launching region.

Similar to the accretion disk, the physical properties of the outflows seem well converged in both higher-resolution simulations and are significantly different from the lowest-resolution simulation. The most striking difference is that only in the higher-resolution simulations (i.e. ,$64\times10$ and $126\times20$), can a very wide, low-velocity tower flow be independently identified. In contrast, in simulation $32\times5$, the magneto-centrifugal jet and the tower flow seem to be merged in the same cells.

Both the momentum as well as the angular momentum content of the outflow shown in panels (a) - (d) of Fig. \ref{fig:conv_outflow} clearly demonstrate that both higher-resolution simulations $64\times10$ and $126\times20$ follow a qualitatively similar evolution. The linear momentum seems to behave similarly for all simulations with a strong initial ejection burst at $4\kyr$, containing a lot of momentum. This very high momentum is primarily carried by entrained envelope material that is moved upwards when the outflow cavity forms but does not necessarily leave the domain.

For $32\times5$ and $64\times10$, the total momentum content of the outflow then decreases until the end of the simulations and another short peak is visible after $20\kyr$. For $126\times20$, the total linear momentum transported by the jet decreases as well after the initial burst but only up to a free-fall time at $52\kyr$. A tower flow is only distinguishable in both higher-resolution simulations $64\times10$ and $126\times20$ shortly after the initial launching of the fast jet component. It becomes dominant both in linear and angular momentum after one free-fall time, when the ram-pressure of the infalling envelope no longer constrains it.

The angular-momentum content of the outflows seems to follow a totally different evolution in the lower-resolution simulation $32\times5$. The reason for this is the impossibility to distinguish between a slow tower flow and the fast magneto-centrifugal jet due to the low resolution.
Comparing this simulation to the jet/wind launching region in the higher-resolution simulations reveals that both processes initially take place so close to each other that they would have to take place in the same cells for simulation $32\times5$. In all simulations, the magnetic field close to the polar axis is strongly dominated by its poloidal components. Therefore, there is no magnetic pressure gradient in the vertical direction that would result in a force comparable to gravity, which would be a good indicator for a magnetic tower flow. Instead, for simulation $32\times5$, the strong centrifugal acceleration close to the polar axis, combined with the large cell size there, leads to a magneto-centrifugal acceleration active in a large volume. At larger radii, a magnetic pressure gradient develops in $32\times5$ as well, not as strong as gravity but still contributing to the vertical acceleration. Therefore, basically the whole outflow in the low-resolution simulation is powered by an unresolved magneto-centrifugal process, aided by a small magnetic pressure gradient, which together accelerate huge amounts of mass with relatively high velocities leading to the momentum-rich outflow visible in Fig. \ref{fig:conv_outflow}.

\begin{figure}
\centering
\begin{subfigure}[b]{0.47\textwidth}
        \includegraphics[width=\textwidth]{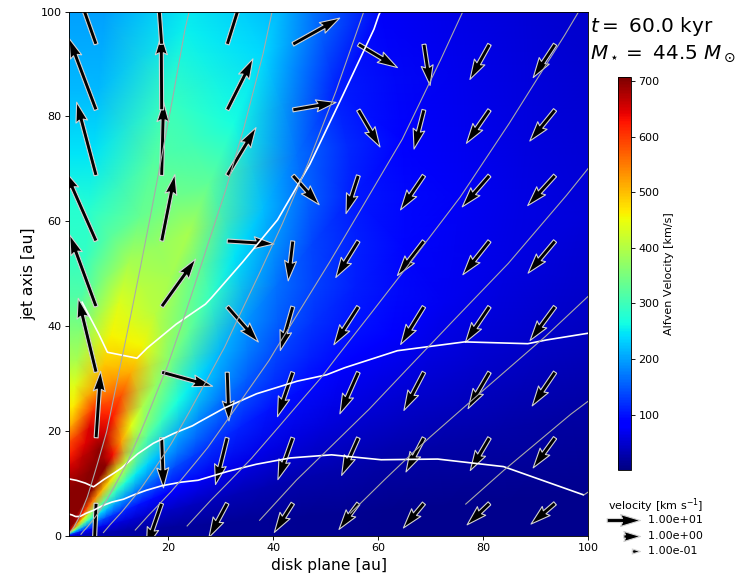}\\
        \caption{} 
\end{subfigure}
\begin{subfigure}[b]{0.52\textwidth}
        \includegraphics[width=\textwidth]{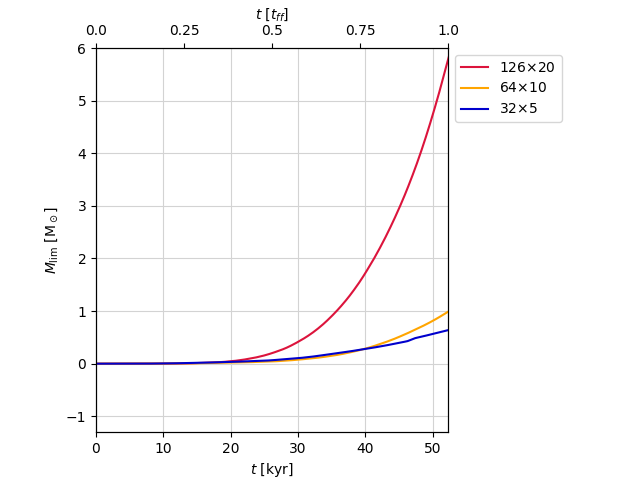}\\
        \caption{} 
\end{subfigure}
\caption{Regions of activity of the Alfv\'en velocity-limiter and the injected mass for different resolutions.
	(a) Alfv\'en velocity after one free-fall time. The white density contours correspond to $10^{-15}\mathrm{g~cm}^{-3}$, $10^{-14}$ g cm$^{-3}$, and $10^{-13}$ g cm$^{-3}$.
	(b) Total mass generated by the Alfv\'en velocity-limiter.
}
\label{fig:conv_star_alfven}
\end{figure}

The last difference we want to point out here is that in simulation $126\times20$, the magneto-centrifugal jet ceases after one free-fall time. Like simulation $126\times20$, simulation $64\times10$ possesses the resolution necessary to develop an independent tower flow. Still, unlike in simulation $126\times20$, the magneto-centrifugal jet is active until the end of the simulation. The reason is that simulation $126\times20$ can resolve the region close to the polar axis with twice the number of cells. This also allows for a higher density contrast, resulting in lower densities directly at the pole. Unfortunately, the polar axis is also the region with the highest magnetic field strengths. Therefore, the Alfv\'en velocity increases drastically, reducing the simulation time step significantly. As described in section \ref{sec:AlfvenLimiter}, we avoid a total halt of the simulation by injecting mass into cells that reach a certain Alfv\'en velocity. This Alfv\'en limiter is implemented in a momentum-conserving fashion, damping the rotational velocity necessary for the centrifugal launching and the resulting flow upwards, eventually halting the fast magneto-centrifugal jet completely. This is also reflected in Fig. \ref{fig:conv_star_alfven}b which shows the mass created by this Alfv\'en velocity-limiter during the full $100\kyr$ simulation.
Panel (a) of the same figure shows the Alfv\'en velocity in the close vicinity of the protostar after one free-fall time, as well as the density contours of its surroundings. As visible there, the Alfv\'en limiter is active in the outflow cavity and can therefore influence the evolution of the jet.
Figure \ref{fig:conv_star_alfven}b shows that the Alfv\'en limiter produces more than $25\Msol$ over the course of the simulation, while simulation $64\times10$ produces less than $5\Msol$. 
In both cases, the bulk of this mass is produced after one free-fall time when the ceasing supply of material leads to very low densities in the polar regions.
This drain of material due to the finite mass reservoir of the initial conditions would most likely not occur if larger-scale mass reservoirs were taken into account; see \citet{Kuiper2018a} for a comparison of these two accretion scenarios.
An Alfv\'en velocity of $700$ km/s is only reached in the cells in vicinity of the polar axis and close to the protostar, as visible in panel (b) for simulation $126\times20$. Therefore, it does not influence the magnetic tower flow at larger radii.
The fact that the fast jet component vanishes in simulation $126\times20$ is not obvious from the linear and angular momentum transported by the jet after one free-fall time. The reason is that the fast jet, although it accelerates gas to very high velocities, only carries relatively low masses compared to the low-velocity tower flow, and therefore also transports much less momentum in total if it is properly resolved.
        
We conducted two further simulations to probe the influences of the Alfven velocity limiter: one with half the maximal Alfven velocity (i.e., $350$ km/s) and one with double the Alfven velocity limit ($1400$ km/s). We found that doubling the maximal Alfv\'en velocity in the domain enables the jet to continue after one free-fall time. However, as expected, this halved the time steps of the integration, after the Alfv\'en velocity reached its limit for the first time. In turn, the simulation with increased maximal Alfv\'en velocity required nearly double the CPU-hours to complete. On the other hand, before one free-fall time, the three simulations appear almost identical. They share the same density, magnetic field, and velocity structure. The only obvious difference is higher jet velocities in simulations with increased Alfven velocity limit. 
        Therefore, we argue that with a similar setup and when increasing the resolution of such a simulation, one has to consider increasing the Alfven velocity limit as well. Here, we do not expect a realistic evolution of the prestellar cloud core after one free-fall time due to the missing radiative feedback. Consequently, a 700km/s Alfven velocity limit provides a reasonable trade-off between computation time and converged behavior before significant differences appear.

Another interesting aspect we can see here is that angular momentum seems to be much more efficiently transported via the magnetic tower than via the fast jet. This is evident in simulations $64\times10$ and $126\times20$, where angular momentum transport increases significantly when the magnetic tower dominates the outflow after one free-fall time.

Concluding, resolutions of $\le 0.17\au$ seem to be sufficient to resolve the magneto-centrifugal jet and a magnetic-pressure-driven tower flow independently. Nevertheless, our simulations with the highest resolution of $0.09\au$ showed a significant influence of the Alfv\'en velocity-limiter that can be avoided with lower resolution. Though this method is widely used and an implementation conserving the momentum is an obvious first choice, other methods of limiting the Alfv\'en velocity should also be considered for future research.

\subsection{The protostar}

\begin{figure*}[h!]
\centering
\begin{subfigure}[b]{0.49\textwidth}
        \includegraphics[width=\textwidth]{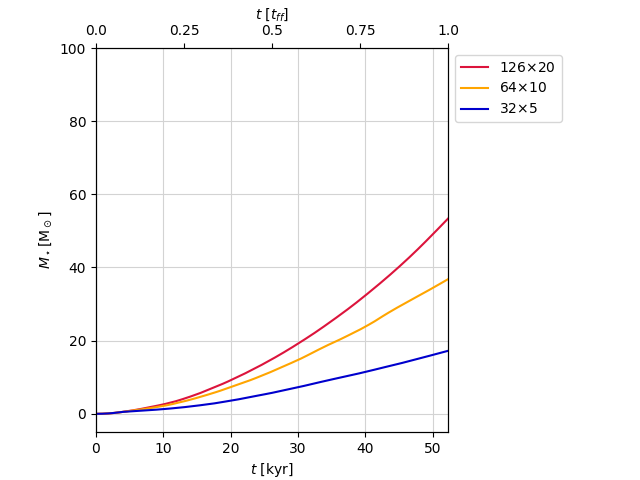}\\
        \caption{} 
\end{subfigure}
\begin{subfigure}[b]{0.49\textwidth}
        \includegraphics[width=\textwidth]{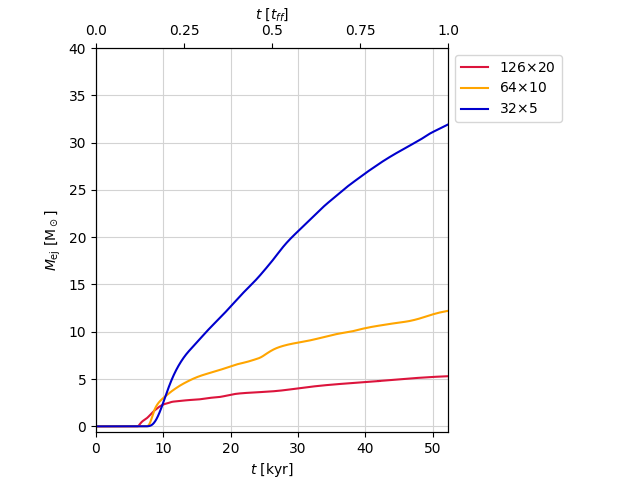}\\
        \caption{} 
\end{subfigure}
\caption{Convergence properties of the accreted and ejected mass for different resolutions.
	(a) Protostellar mass as a function of time.
	(b) Mass ejected from the computational domain as a function of time.
}
\label{fig:conv_star}
\end{figure*}

The protostellar mass depicted in Fig. \ref{fig:conv_star}a shows that the simulations are clearly resolution dependent with respect to the protostellar evolution. These results partly mirror the insights that we gained from Sect. \ref{sec:conv_outflows_res}. For lower resolutions, a significantly higher mass is ejected, as visible in Fig. \ref{fig:conv_star}b, and this ejection, by necessity, robs mass from the protostar.
\begin{figure}[h!]
\centering
\includegraphics[width=0.5\textwidth]{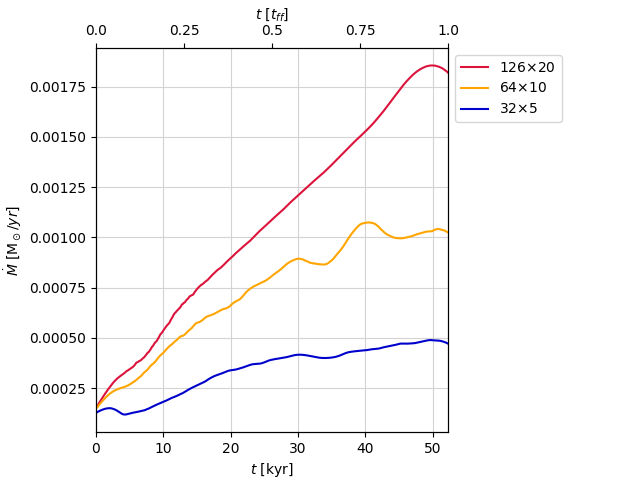}\\
\caption{Accretion rate to the protostar as a function of time, and for different resolutions.}
\label{fig:conv_star_accr}
\end{figure}
This is also reflected in the accretion rate to the protostar visible in Fig. \ref{fig:conv_star_accr}.

To take this into perspective: for simulation $126\times20$ by one free-fall time $50\Msol$ has been accreted to the protostar and only $\simeq 5\Msol$ has been ejected. 
For $32\times10$, in contrast, the poorly resolved magneto-centrifugal launching mechanism ejects vast amounts of mass, such that at one free-fall time $32\Msol$ has already been ejected from the simulation, while only $\simeq 18\Msol$ has been accreted. 

As a conclusion, different resolutions have a very significant influence on the outcome of such collapse simulations. While the disk evolution is not strongly affected, properly resolving magneto-centrifugal jets is of major importance, and the simulation with $32\times5$ cells, corresponding to a resolution of $0.37\au$ at the protostar, is not able to achieve this.
We deem both the fact that the magneto-centrifugal jet ceases in the highest resolved simulation as well as the unrealistically low ejection-to-accretion ratios to be the result of the Alfv\'en limiter; without it, however, a simulation with such a long temporal span and with such great resolutions would not have been possible. Also, the type of Alfv\'en limiter we implemented is the most commonly used, as other methods of changing the Alfv\'en velocity often have much greater complications.
It is worth noting here that even before reaching one free-fall time, radiative effects become essential for the consequent realistic evolution of the system. We expect them to have an impact on the bipolar low-density cavity region when the protostar reaches $\ge 20\Msol$ which happens at $\simeq30\kyr$ in our simulations.
Here, we focus on the convergence properties of the MHD physics and leave the interaction with the radiation feedback, as modeled in \citet{Kuiper2018a} for example, to future studies.

Up until this point, though, we consider our simulations to be a realistic approximation of reality due to their very high spatial resolution and due to the good agreement of the higher-resolution simulations, especially concerning the launching event and the consequent evolution of the accretion disk and both types of outflow.
Even though we do not expect a realistic outcome for our simulations after one free-fall time, all the described effects show the importance of thorough convergence analyses. Another important numerical parameter of our simulations is the sink-cell size. Its influence on the results is analyzed in the following convergence study.

\section{Convergence properties for different sink-cell sizes}
\label{Convergence_sinks}

As described in Sect. \ref{sec:Grid}, we use a spherical grid with cell sizes that increase logarithmically in the radial direction. This greatly increases the resolution in the center and close to the protostar, while we can still cover a large cloud core up to $0.1\pc$. However, smaller cell sizes also result in small computational time steps and, therefore, we have to truncate this grid close to the origin. There we introduce a sink cell that models the forming protostar and the innermost regions around it.\footnote{More or less all grid-based astrophysical simulations also use the concept of sink cells or sink particles, because with increasing densities, at some point, they can no longer resolve physically important scales with enough cells (usually the Jeans length or Toomre $Q$) and at that point it becomes more useful to use a subgrid model to simulate the insides of such a cell.}

In the following sections, we discuss the effects of changing the size of this sink cell. Here, we compare four simulations with sink-cell sizes of $30.3$, $10.9$, $3.1$, and $1.0$ au. Again, all other physical and numerical parameters are kept. However, instead of using the same number of cells in each direction, we keep the resolution and with it the cell sizes and positions constant. This means that the grid with a $30\au$ sink cell has a lower number of cells in the radial direction than a grid with a smaller sink radius. Still, the cells at radii greater than $30\au$ have the exact same positions and sizes in each direction. This way, a grid with smaller sink radius simply has additional cells closer to the forming protostar. The following sections discuss the influence of sink-cell size on the  evolution of the disk, that of the outflows, and that of the protostar. 

\subsection{The accretion disk}

\begin{figure*}[h]
\centering
\begin{subfigure}[b]{0.49\textwidth}
        \includegraphics[width=\textwidth]{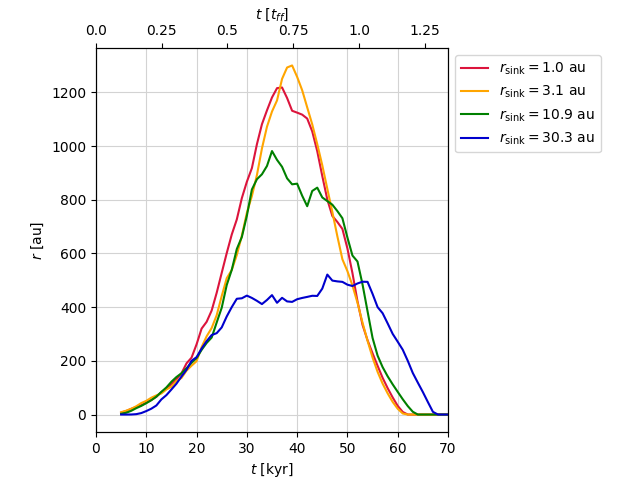}\\
        \caption{} 
\end{subfigure}
\begin{subfigure}[b]{0.49\textwidth}
        \includegraphics[width=\textwidth]{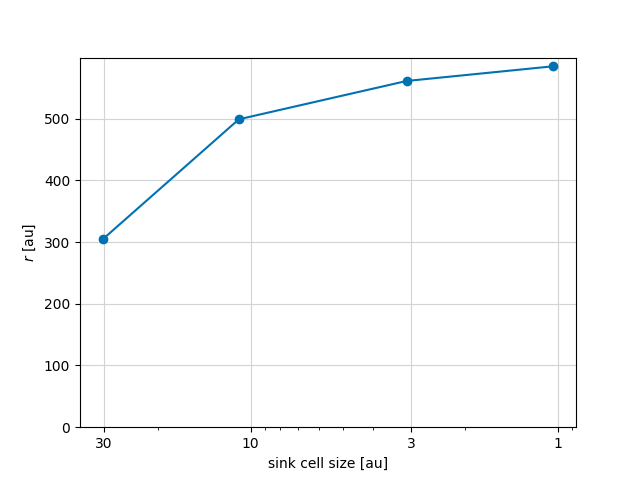}\\
        \caption{} 
\end{subfigure}
\caption{Convergence properties of the disk radius for different sink-cell sizes.
	(a) Disk radius as a function of time.
	(b) Time-averaged disk radius as a function of sink-cell size.
}
\label{fig:conv_disk_sinks}
\end{figure*}

The sink-cell size impacts the convergence properties of the disk to a lesser degree than resolution changes. The disk radius closely follows a similar evolutionary trend regardless of sink-cell size as visible in panel (a) of Fig. \ref{fig:conv_disk_sinks}. Again, the disk lifetime seems to be mostly unaffected, though there are major discrepancies between simulations with very large and small sink cells, which is reflected by their corresponding time-average plotted in panel (b). 
The simulation with the largest sink-cell radius $\rsink=30.3\au$ shows a very significant divergence as its disk radius does not reach the same global maximum of $\simeq 1200\au$ in contrast to the simulations with $\rsink=1.0\au$ and $\rsink=3.1\au$, but remains only below $600\au$ at all times. The simulation with $\rsink=10.9\au$ shows intermediate values for the disk radius.

Further analysis shows that the part of the accretion disk that lies in the domain has a relatively similar-shaped density structure, but differs in angular-momentum content. For the simulation with larger sink-cell radii $\rsink=30.3\au$ and $\rsink=10.9\au$, this results in a weaker centrifugal support.
This probably stems from the fact that more material and with it more angular momentum was accreted into the larger sink cell and is not available to be redistributed in the disk to establish the necessary velocities.

As visible in Fig. \ref{fig:conv_disk_sinks}b, with decreasing sink-cell size our results clearly approach a converged state with very insignificant differences between $\rsink=3.1\au$ and $\rsink=1.0\au$. As a conclusion, a sink radius of $3.1\au$ is sufficient to capture the physics of the accretion disk realistically. In contrast, already a sink radius of $10.9\au$, which is a commonly used size for a sink particle/sink cell, has problems with the realistic accretion of angular momentum that is then missing from the forming disk.

\subsection{The magneto-centrifugal jet and the magnetic tower flow}

\begin{figure*}[h]
\centering
\begin{subfigure}[b]{0.49\textwidth}
        \includegraphics[width=\textwidth]{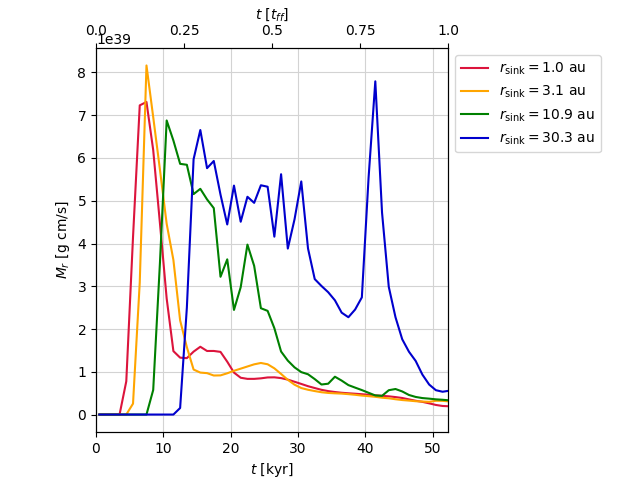}\\
        \caption{} 
\end{subfigure}
\begin{subfigure}[b]{0.49\textwidth}
        \includegraphics[width=\textwidth]{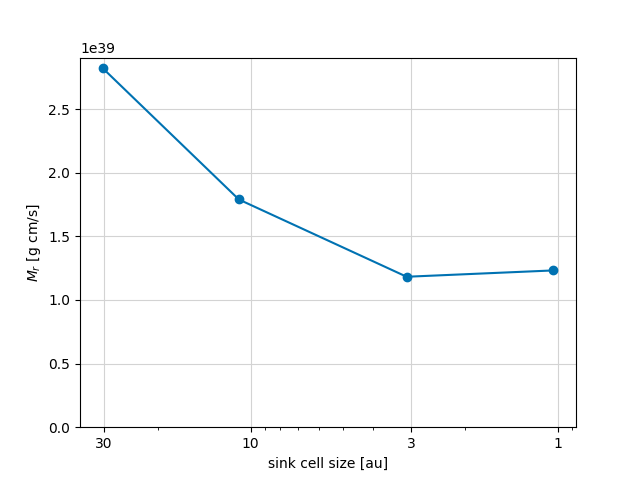}\\
        \caption{} 
\end{subfigure}
\begin{subfigure}[b]{0.49\textwidth}
        \includegraphics[width=\textwidth]{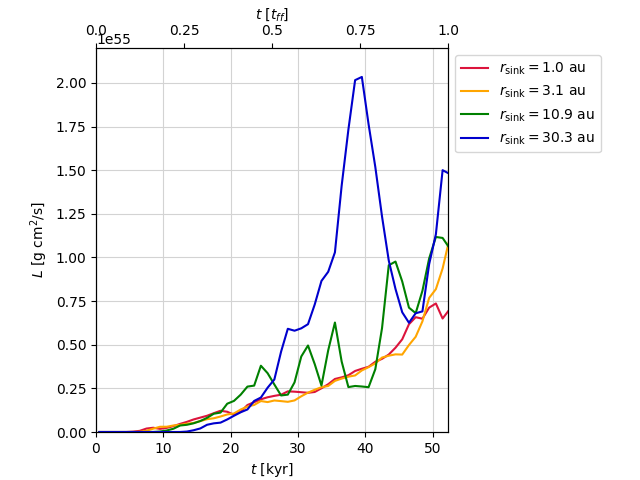}\\
        \caption{} 
\end{subfigure}
\begin{subfigure}[b]{0.49\textwidth}
        \includegraphics[width=\textwidth]{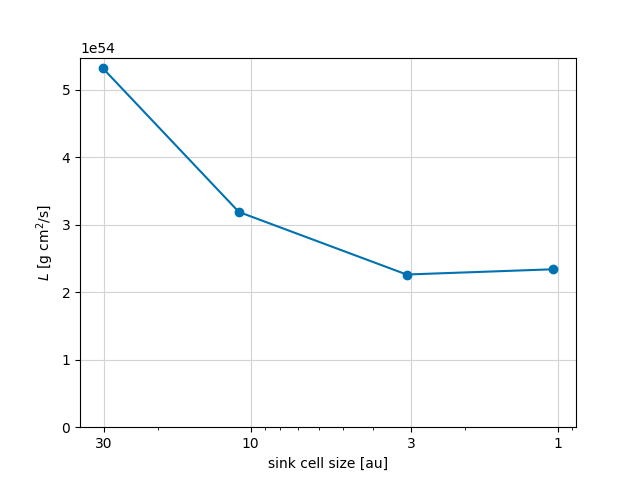}\\
        \caption{} 
\end{subfigure}
\caption{Convergence properties of the outflow for different sink-cell sizes.
	(a) Outflow momentum as a function of time.
	(b) Time-averaged outflow momentum as a function of sink-cell size.
	(c) Outflow angular momentum as a function of time.
	(d) Time-averaged outflow angular momentum as a function of sink-cell size.
}
\label{fig:conv_outflow_sink}
\end{figure*}

The outflows behave similarly to the disk with varying sink-cell sizes. Here, we can clearly distinguish the evolutionary tracks of the simulation with $r_\textrm{sink} = 30\au$ from simulations with $r_\textrm{sink} = 3\au$ and $r_\textrm{sink} = 1.0\au$, while the simulation with $r_\textrm{sink} = 10\au$, again, shows an intermediate behavior.

As visible in panel (a) of Fig. \ref{fig:conv_outflow_sink}, showing the total linear outflow momentum as a function of time and panel (b) showing its time-average over the course of the simulation, the simulation with $r_\textrm{sink} = 30.3\au$ launches its outflow far later than the other simulations but with a much higher momentum over a long period of time, also resulting in much higher average momenta. Generally, lower sink-cell radii tend to produce earlier outflows; simulation $\rsink=1.0\au$ launches its outflow at $4\kyr$, $\rsink=3.1\au$ at $5\kyr$, $\rsink=10.9\au$ at $8\kyr$, and $\rsink=30.3\au$ at $12\kyr$.
Simulations with $r_\textrm{sink} = 3\au$ and $r_\textrm{sink} = 1.0\au$, on the other hand, show a strong but short-lived outflow and then immediately settle down to lower average outflow momenta.
This effect is foremost due to a broader launching region above the sink cell for larger sink radii which is due to two contributing factors.
First, the physical extent of the cells launching the outflows increases with sink-cell size, since the outflows are always launched in close proximity to the protostar and, consequently, launch very close to the sink cells, resulting in larger launching radii for simulations with larger sink-cell sizes.
Second, the larger sink-cell sizes also result in slightly different magnetic field morphologies, effectively allowing even more cells to launch an outflow.
This ultimately leads to an acceleration that affects a significantly larger volume, in turn launching more mass.

As also mentioned in Sect. \ref{Convergence_res}, we found that most of the angular momentum is transported by the tower flow instead of the magneto-centrifugal jet and that the tower flow is active at larger radii. Therefore, it is not surprising that the angular momentum transport through outflows does not differ significantly between simulations with different sink cell radii. This is very apparent in Fig. \ref{fig:conv_outflow_sink}c, showing the angular momentum content of the outflow as a function of time, and in Fig. \ref{fig:conv_outflow_sink}d, showing its time average, as the differences in angular momentum in the outflow are much smaller than the differences we just observed for the linear momentum.

\begin{figure}
\centering
\includegraphics[width=0.5\textwidth]{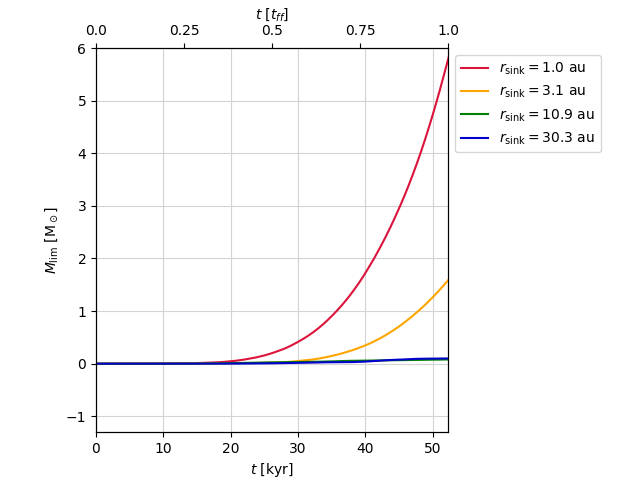}
\caption{Total mass generated by the Alfv\'en velocity-limiter for different sink cell radii.}
\label{fig:conv_star_alfven_sinks}
\end{figure}

After one free-fall time, all simulations produce an extremely wide outflow tower flow. However, there is a surprising difference. In the simulation with $\rsink=30.3\au$, the linear and angular momentum in the outflow increases much less drastically than in the other simulations. This reveals another interesting aspect of this tower flow: it is extremely sensitive to changing magnetic field topology.
In the simulation with $\rsink=30.3\au$, the wider and continuously active magneto-centrifugal outflow launched close to the large sink cell disturbs the magnetic field morphology at larger distances from the protostar than it does for lower sink cell radii. Therefore, while in simulations with small sink-cell sizes it slowly engulfs the whole region from a few $100\au$ up to $>10000\au$, the relatively weak tower flow can only develop in a few parts of the simulation for larger sink cells.

        The mass created by the Alfv\'en velocity limiter is naturally higher in simulations with smaller sink cells, as closer to the origin the highest magnetic field strengths are reached. However, as visible in Fig. \ref{fig:conv_star_alfven_sinks}, it produces only a relatively small mass before one free-fall time, for larger sink-cell sizes only a fraction of a solar mass is produced up to $52\kyr$. Additionally, in all simulations it is only active directly above the protostar and not in extended regions above the disk. Therefore, it is probably not responsible for disturbing the magnetic field topology on larger scales in simulations with larger sink cells.

Similarly, the fast jet component totally vanishes for the simulations with $r_\textrm{sink} = 3.1\au$ and $r_\textrm{sink} = 1.0\au$ which is again due to the small cell sizes in the polar jet launching region where the Alv\'en limiter produces the most mass. It also suppresses the acceleration due to angular momentum conservation, analogous to simulations with high resolutions discussed in Sect.  \ref{sec:conv_outflows_res}. This time, however, the resolution is constant in each simulation, but the cell sizes reduce close to the origin. 

Generally, the influences of the sink-cell size are most apparent closer to the origin and seem to disappear for larger radii at least with respect to the ejection-to-accretion efficiency, as indicated in
panels (a) and (c) of Fig. \ref{fig:conv_star_sinks}, when looking at the total sum of material ejected from the domain and comparing it to the total accreted mass. There, we find a ratio of $12 - 22 \%$ over the total simulation time, also for smaller sink-cell sizes, such that the accretion-to-ejection efficiency on large spatial and temporal scales is of the expected magnitude.

Our discussion shows that sinks with $3\au$ and less are required for a converged result, and most (if not all) Cartesian AMR simulations use sink particles  with an accretion
radius of four times the highest grid resolution, that is, much larger than $3\au$, in many cases even larger than $30\au$. 
The most similar study that combines AMR with sink cell creation is \citet{Seifried2011} with a $12.6\au$ sink cell and they argue that their outflow-launching mechanism shows contributions from magnetic pressure as well as magneto-centrifugal launching. \citet{Peters2011} use AMR with sink cells of $590$ au in accretion radius and they find a relatively massive, slow, pressure-driven wind in their simulations. Other authors combine nested grids with a sink cell treatment. For example, \citet{Machida2014} use this kind of grid to analyze the dependence of disk formation on sink-cell size . Unfortunately, they do not explicitly discuss their influences on outflow launching, even though they use small sink cells of radii down to $1\au$.

\subsection{The protostar}

\begin{figure*}[h]
\centering
\begin{subfigure}[b]{0.49\textwidth}
        \includegraphics[width=\textwidth]{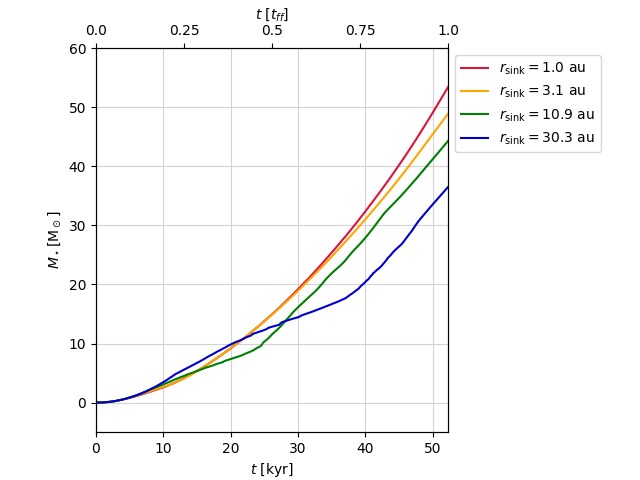}\\
        \caption{} 
\end{subfigure}
\begin{subfigure}[b]{0.49\textwidth}
        \includegraphics[width=\textwidth]{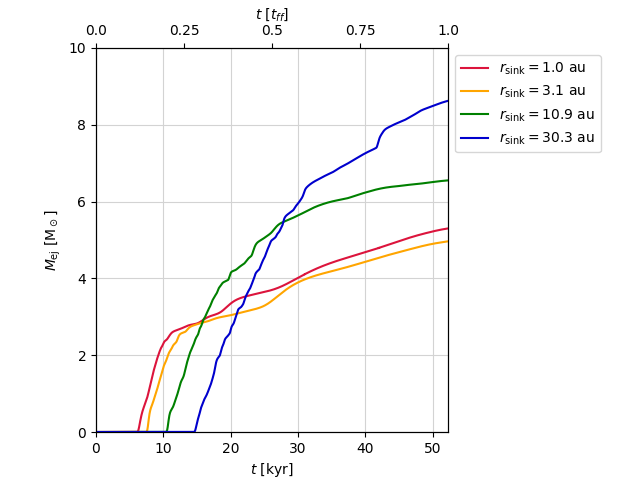}\\
        \caption{} 
\end{subfigure}
\caption{Convergence properties of the accreted and ejected mass for different sink-cell sizes.
	(a) Protostellar mass as a function of time.
	(b) Mass ejected from the computational domain as a function of time.
}
\label{fig:conv_star_sinks}
\end{figure*}

Comparing the convergence of simulations with different sink radii with respect to accretion to the protostar reveals a nonintuitive feature: simulations with larger sink radii accrete less to the protostar and eject more mass out of the domain, as visible in Fig. \ref{fig:conv_star_sinks}a, showing the protostellar mass as a function of time and Fig. \ref{fig:conv_star_sinks}b showing the mass ejected from the simulation. Intuitively, one would expect that a larger sink would ultimately accrete material earlier, before it can be ejected by magneto-centrifugal ejection and, therefore, should show the opposite behavior.

This feature is a consequence of two processes discussed above. Firstly, simulations with larger sinks produce broader magneto-centrifugal jets that also hinder the accretion to the protostar at intermediate polar angles between $30\degree$ and $70\degree$ and only allow accretion close to the equatorial plane, where centrifugal forces and magnetic pressure stabilize the accretion disk further, giving the outflows more time to eject mass. Secondly, they generally produce more massive outflows, reducing the mass available for accretion.

\begin{figure}[h!]
\centering
\includegraphics[width=0.5\textwidth]{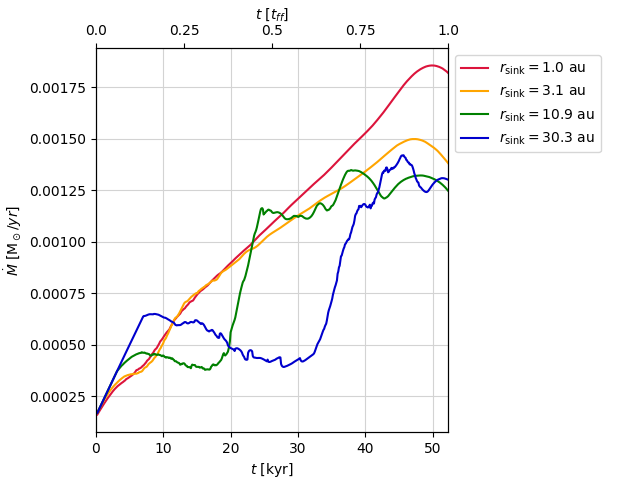}\\
\caption{Accretion rate to the protostar as a function of time, and for different sink-cell sizes.}
\label{fig:conv_star_accr_sinks}
\end{figure}

The first point is especially apparent in Fig. \ref{fig:conv_star_accr_sinks}, showing the accretion rate to the protostar as a function of time, where the accretion after the initial jet launching-event at $4-12\kyr$ (depending on the sink-cell size) is further postponed for simulations with larger sink radii ($\rsink>3.1\au$). 
The ejected mass shown in Fig. \ref{fig:conv_star_sinks}b shows a similar picture. For larger sink cell radii, the outflows are launched at a later time, accelerate more mass and, consequently, eject more mass out of the simulation, which is then not available for the accretion to the protostar either.

Concluding, we find that the influence of varying sink-cell size is similar to the influences of changing resolution, while naively, one would expect that a smaller sink radius allows the jet-launching region to extend closer to the central host star, that is, deeper into the potential well. This, in turn, could lead to a faster magneto-centrifugal jet and, as a consequence, we would also have expected to see an increase in the momentum transported by it.
The opposite results are visible here: while larger sink-cell sizes lead to a later launching of magneto-centrifugal jets, they are loaded with more mass and their wider launching region hinders the accretion to the protostar, resulting in lower protostellar masses and higher ejection-to-outflow efficiencies. Further, the influences of the sink radii seem to reduce with distance from the central object. We also find that a sink cell radius of $3.1\au$ seems to show all the qualitative behavior we have seen for lower sink radii, while a simulation with $\rsink = 10.9\au$ still showed major differences, especially with respect to outflow properties. Therefore, to resolve the physics of the outflows properly, we deem a sink-cell size of $\leq 3.1\au$ necessary.

\begin{figure}[h!]
\centering
\includegraphics[width=0.5\textwidth]{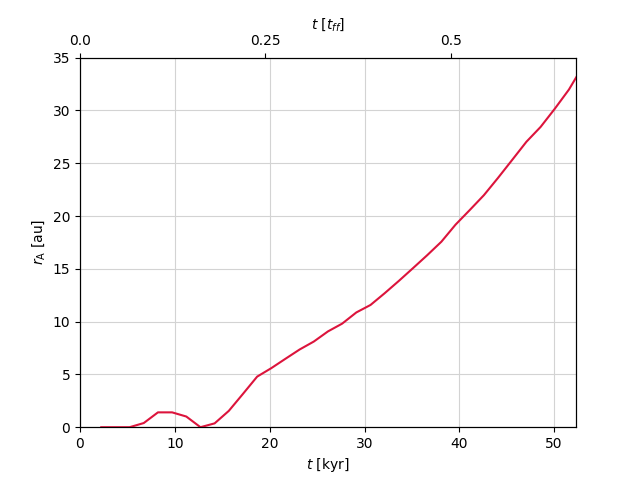}\\
\caption{Alfv\'en radius in the equatorial plane as a function of time in the simulation with $\rsink = 1.0\au$.}
\label{fig:alfven_radius_sinks}
\end{figure}

This necessary sink radius is tied to the physics of magnetic jet collimation. If the jet collimation region is not included in the computational domain it enables the jet to entrain material from an unrealistically large region. We find that the anchoring points of the magnetic field lines that are responsible for magnetic collimation are essential to be included in the simulation as they provide an important contribution to the magnetic pressure constraining the jet. More specifically, the Alfv\'en radius in the accretion disk has to be included in the simulation, that is, outside of the sink cell to result in a realistic jet collimation, avoiding artificial entrainment.
We can track this result in Fig. \ref{fig:alfven_radius_sinks}, showing the Alfv\'en radius in the equatorial plane as a function of time in simulation $\rsink=1.0\au$. Here, the timescale when the Alfv\'en radius reaches the corresponding sink cell radii, that is, at $3.1\au$,  $10.9\au$, and $30.3\au$, the outflow (linear and angular) momentum of simulations with larger sink cells approaches the values of the simulation with $\rsink=1.0\au$. 

Equipped with the knowledge of our convergence study, we are able to assess the reliability of the physical results of our simulations. Well converged results are presented in the following section, before the whole project is summed up.

\section{Physical results}
\label{Physical_Effects}

In this section, we analyze the data from the simulation with $126\times20$ and $64\times10$ cells and with a sink-cell radius of $\rsink=1.0\au,$ that is, the highest- and second-highest-resolution simulation with the smallest sink cell.
As shown in the previous sections, the disk physics as well as both launching mechanisms are soundly resolved in both simulation and promise realistic results at least up to one free-fall time. After this, the simulation clearly shows signs of unrealistic behavior due to missing physical effects. The highest-resolution simulation is best suited for the analysis of all but one aspect of the physical processes: the magneto-centrifugal jet. As the previous discussion showed, the Alfv\'en limiter significantly influences the magneto-centrifugal jet in the highest resolution, while other aspects, like the launching process and the independently resolved magneto-centrifugal jet and tower flow, are converged in both simulations.
We begin our discussion, in the following subsections, with the physics during the initial free-fall epoch, continuing with jet launching, disk formation, and the formation of the wide-angle wind in detail, considering all the forces involved, namely gravity, centrifugal, and magnetic forces.

\subsection{Gravitational infall and disk formation}

Our simulations begin at the onset of gravitational collapse. Consequently, gravity dominates the whole mass reservoir and the prestellar core collapses near free fall for the first $4\kyr$.
Angular momentum conservation leads to a notably rotationally flattened envelope, more pronounced in the very center. Magnetic fields are of minor importance here, since the initial cloud has only a weak magnetization with average magnetic field strength up to $ 4\times 10 ^{-5}\Gs$ in the innermost $1000\au$.

\begin{figure}
\centering
\includegraphics[width=0.49\textwidth]{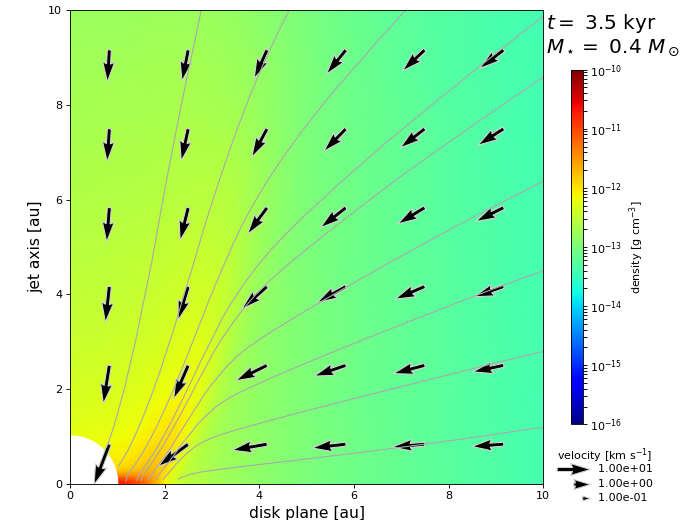}
\caption{Density structure at the onset of disk formation, overlaid with magnetic field lines.}
  \label{fig:disk_emerging}
\end{figure}

\begin{figure*}[h!]
\centering
\begin{subfigure}[b]{0.75\textwidth}
        \includegraphics[width=0.835\textwidth]{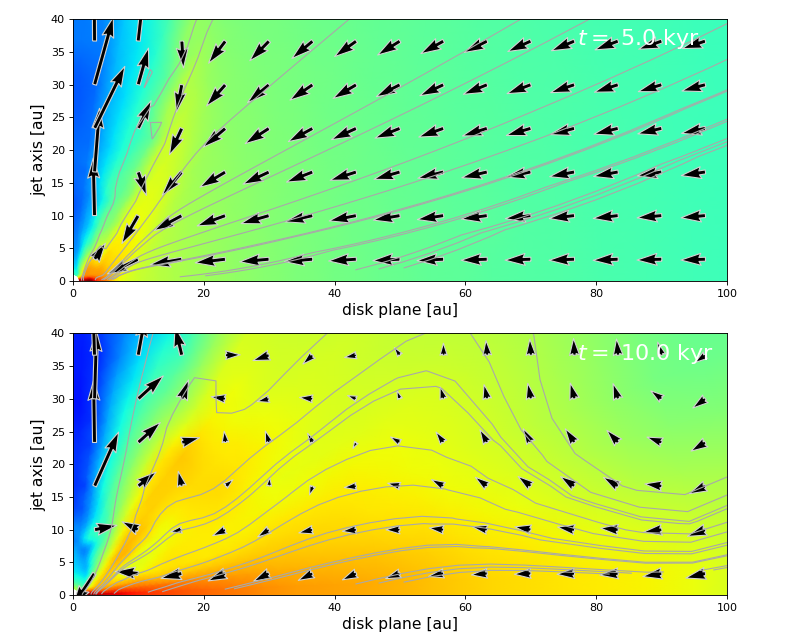}
        \includegraphics[width=0.155\textwidth]{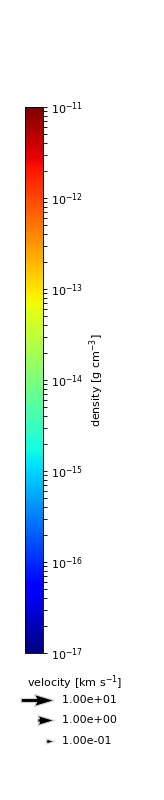}
        \caption{} 
\end{subfigure}
\begin{subfigure}[b]{0.75\textwidth}
        \includegraphics[width=0.835\textwidth]{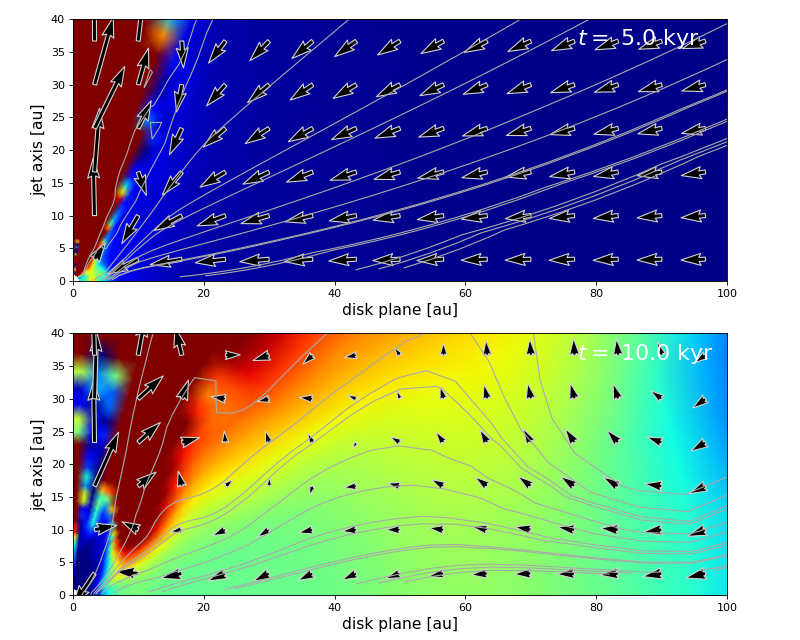}
        \includegraphics[width=0.155\textwidth]{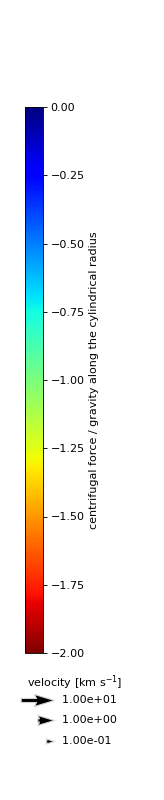}
        \caption{} 
\end{subfigure}
\caption{Build-up of centrifugal support in the disk. 
	(a) Density structure at $5\kyr$ (upper panel) and at $10\kyr$ (lower panel).
	(b) Centrifugal force divided by gravity at $5\kyr$ (upper panel) and at $10\kyr$ (lower panel).
}
\label{fig:disk_formation}
\end{figure*}

From here on, the first disk-like structure develops. The material is continuously accreted from the envelope and settles down in the midplane. While this disk-like structure is clearly visible in the density plot in Fig. \ref{fig:disk_emerging}, it is not yet stabilized, either by centrifugal or by magnetic forces.
Even though it still has a significant infall velocity and is constantly accreted, it is visible for an extended amount of time because it is fed by the infalling envelope, meaning that the material is always replenished.
The subsequent formation process of the disk is depicted in Fig. \ref{fig:disk_formation}; panel (a) shows the density structure and panel (b) shows the centrifugal force compared with gravity at $5\kyr$ (on the top) and $10\kyr$ (on the bottom), respectively.
The colors are chosen such that shades of blue indicate that gravity is stronger than the centrifugal force, shades of green mean that a region is close to force equilibrium, and shades of yellow and red visualize the predominance of centrifugal forces, leading to an acceleration towards greater radii.

At the shown time of $5\kyr$, the jet has already launched and matured into a stable state and the disk has established a centrifugal support up to $\simeq 5\au$. Magnetic fields only play a role in the magneto-centrifugal launching process of the jet close to the polar axis and do not support the disk here to an important degree, as the Lorentz force in radial direction, and in the innermost $\simeq 5\au$ of the disk, only accounts for a few percent of the centrifugal force.
During the subsequent $5\kyr$, the disk expands outwards, also expanding its radius of centrifugal support. At $10\kyr$, a centrifugally supported high-density disk structure is already clearly visible in the lower panels of Fig. \ref{fig:disk_formation}, which effectively extends from the protostar to $\simeq100\au$. Also visible here is that the innermost $5\au$ of the disk are not supported by centrifugal forces. Instead, there, the increasing magnetic pressure has reached a magnitude to rival gravity, thereby supporting the accretion disk.

\begin{figure*}
\centering
\begin{subfigure}[b]{0.49\textwidth}
        \includegraphics[width=\textwidth]{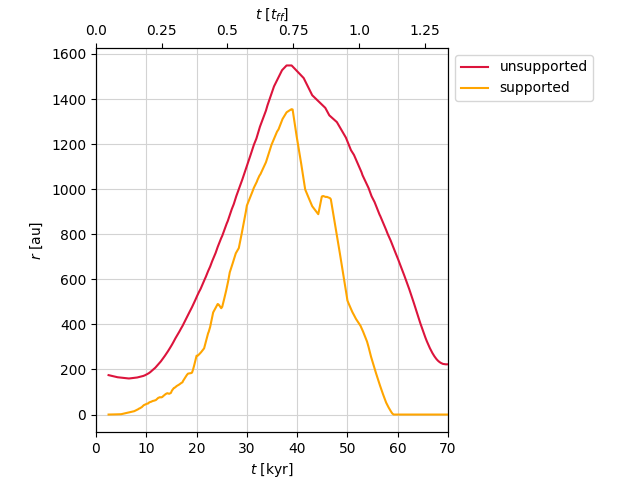}
        \caption{} 
\end{subfigure}
\begin{subfigure}[b]{0.49\textwidth}
        \includegraphics[width=\textwidth]{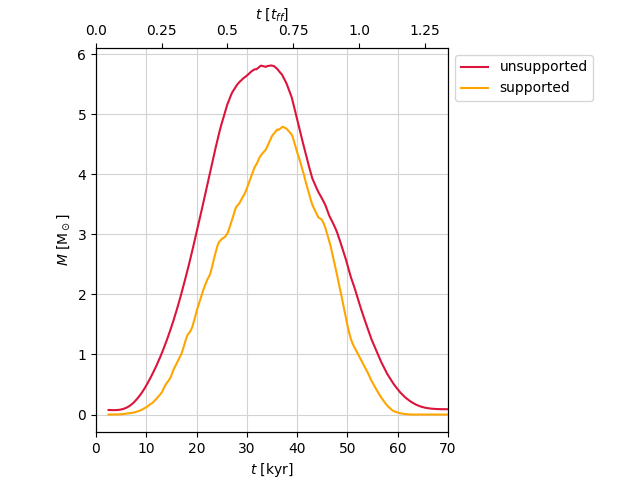}
        \caption{} 
\end{subfigure}
\caption{Time evolution of the disk: The red and yellow lines correspond to the unsupported and centrifugally supported disk, respectively.
	(a) Disk radius as a function of time.
	(b) Disk mass as a function of time.
}
\label{fig:disk_time_evo}
\end{figure*}

The subsequent temporal evolution of the disk is depicted in Fig. \ref{fig:disk_time_evo}. Here, we defined a threshold density of $10^{-15}$ g cm$^{-3}$ to decide whether a region belongs to the disk or to the envelope. Additionally, we consider centrifugal support; we consider a region centrifugally supported if the magnitude of the centrifugal force lies within $\pm 5\%$ of the gravitational force.
From panel (a) of Fig. \ref{fig:disk_time_evo} we can see that a part of the disk (on the order of $200\au$) is not centrifugally supported, although it reaches high densities of $10^{-15}$ g cm$^{-3}$, and that this unsupported part of the disk accounts for a relevant part of the mass of the disk, as visible in panel (b) of the same figure.

In principal, there are two parts of the disk that are not supported by centrifugal forces. The innermost part of the disk, close to the protostar, is primarily magnetically dominated. Here, the disk is supported by the magnetic pressure in the equatorial plane. This part accounts for the first $\simeq5\au$ of the pseudo-disk at $10\kyr$ and does not contribute to the difference in disk radius, since we plot the maximal radius of the centrifugal disk here, from the protostar to its outer rim.
In this part of the disk, the magnetic field is primarily pointing upwards from the equatorial plane. Therefore, angular momentum transport by the magnetic field in the disk is inefficient and its transport takes place primarily by viscous forces.
At the same time, the high densities in this area will shield this region from ultraviolet radiation, resulting in low ionizations and high magnetic diffusivities. We take this into account in our simulations by increasing magnetic diffusivity with density, resulting in particularly high values of diffusivity in this region close to the protostar.
Therefore, while the magnetic pressure in this region is high, the viscously decelerated material can still be accreted efficiently, and the region is not perfectly supported against gravity.
In such accretion disks around forming high-mass stars, angular momentum transport is expected to be primarily due to gravitational torques of the forming spiral arms; see \citet{Kuiper2011} and \citet{Meyer2017, Meyer2018} for details.

The larger unsupported part belongs to the outer rim of the disk, where material is accreted from larger radii and has not yet spun up to Keplerian velocities but will eventually become part of the centrifugally supported region.
This outcome on the physics at the envelope-disk transition is also in agreement with previous simulations of massive disk formation which did not take into account magnetic fields \citep{Kuiper2011, Kuiper2018a} and again shows that only the innermost part of the disk can be supported by magnetic forces.

\begin{figure*}
\centering
\includegraphics[width=\textwidth]{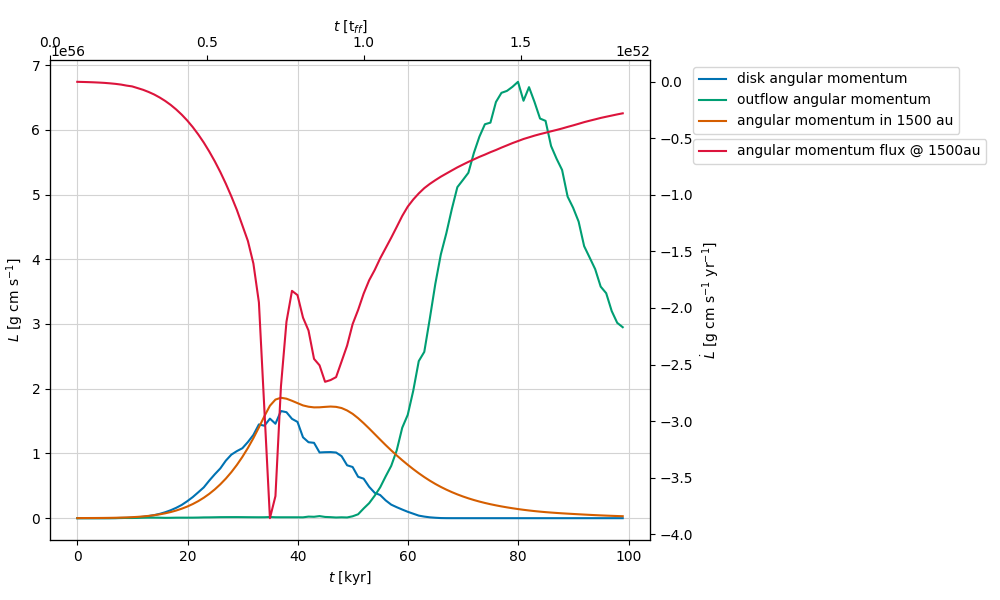}
\caption{Angular momentum in the disk, the outflow, and a sphere of $1500$ au radius around the protostar, as well as the angular momentum flux through the surface of the sphere as a function of time.}
\label{fig:ang_mom_outflow_disk}
\end{figure*}

The envelope is not only a mass reservoir but also a source of angular momentum. Therefore, when less material from larger radii rains onto the disk, it cannot replenish the angular momentum lost by magnetic breaking, viscosity and potentially by outflows. Therefore, the centrifugal support of the disk vanishes after one free-fall time, as visible in Fig. \ref{fig:disk_time_evo}. Additionally, the missing ram pressure exerted by the envelope then enables a much larger tower flow to develop that efficiently removes angular momentum from the disk, contributing heavily to its demise.

This effect can be visualized by comparing the green line in Fig. \ref{fig:ang_mom_outflow_disk}, which shows the total angular momentum in the outflow, with the angular momentum content in the disk in blue. Here, we consider all regions with positive vertical velocities as part of the outflows, such that they include the magneto-centrifugal jet and the tower flow. Additionally, Fig. \ref{fig:ang_mom_outflow_disk} shows the angular momentum within a sphere of $1500\au$ radius (in brown) and its net flux though its surface, which we expect to be dominated by the angular momentum loss of the disk.
The correlation of angular momentum in the disk and the angular momentum flux is obvious: during the whole simulation, the net flux is negative. Therefore, more angular momentum is accreted into $1500\au$ than is ejected which is also reflected by the angular momentum content of the disk. Not visible here is the amount of angular momentum extracted by viscous forces and magnetic breaking that leads to the constant reduction of angular momentum in the disk. 

The maximal influx of angular momentum to the disk is reached at $35\kyr$, corresponding to the maximum angular momentum content of the disk. After that, viscosity and magnetic breaking reduce the angular momentum in the disk faster than the envelope replenishes it. From the angular momentum content of the outflow we can deduce that only an insignificant part of the angular momentum extraction is carried out by outflows before one free-fall time, as the magnetic tower is not able to lift material up against the ram-pressure of the infall, and the magneto-centrifugal jet is launched from low radii, where it has a small magnetic lever arm and can only extract an insignificant portion of angular momentum.
However, after one free-fall time at $52\kyr$, the outflow begins to carry a massive part of the total angular momentum due to the developing tower flow.
\begin{figure*}
\centering
        \includegraphics[width=0.835\textwidth]{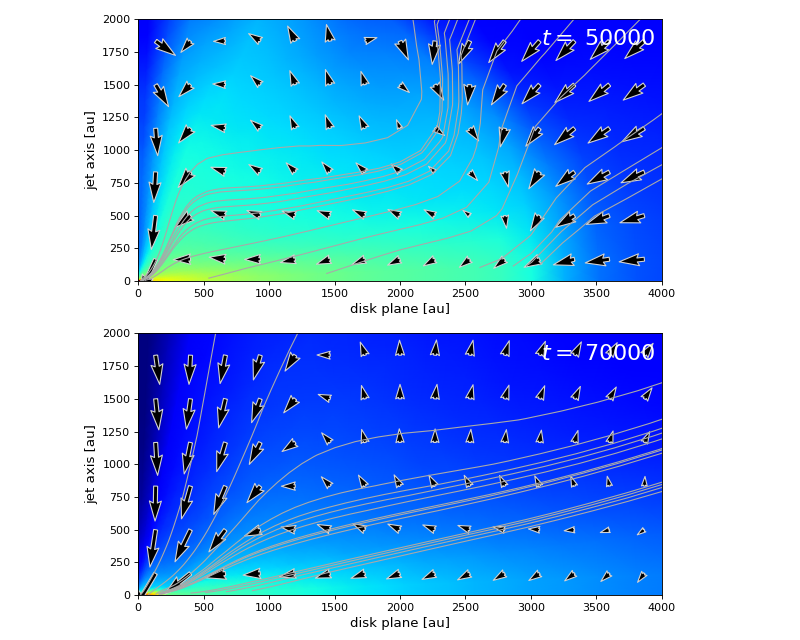}
        \includegraphics[width=0.155\textwidth]{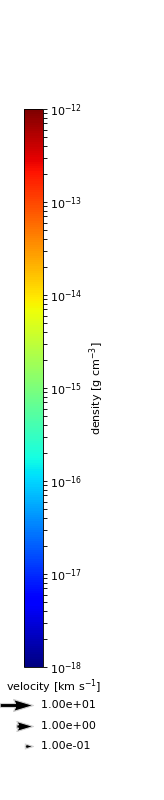}
        \caption{Density structure of the accretion disk at $50\kyr$ (upper panel) and at $70\kyr$ (lower panel).
}
\label{fig:acc_clump_after_fft}
\end{figure*}

Another process occurring after one free-fall-time is that the formerly high scale height of the disk decreases constantly. This is a result of the compression of the magnetic field lines towards the disk and closer to the midplane, as visible when comparing the upper panel of Fig. \ref{fig:acc_clump_after_fft} (showing the situation at $5\kyr$) with its lower panel (showing the situation after envelope dissipation at $70\kyr$). 
        As a result, the magnetic field lines that are now less inclined with respect to the disk can much more efficiently couple particles differentially rotating at lower radii with particles at higher radii, potentially increasing the angular momentum transport within the accretion disk. On the other hand, the vanishing envelope reduces the efficiency of magnetic breaking at higher radii of the disk.
        The support by the outer envelope and its constant influx of material, which was previously leading to an increased density above the disk, also reduces such that the material in the disk's atmosphere can be accreted faster than it is replenished ultimately removing this region of increased density. 

As we have seen, the evolution of the outflows has a prominent influence on the disk formation, but its physical evolution is an important topic in itself and is analyzed in the following subsection.

\subsection{The magneto-centrifugal jet}
\label{sec:jet}

The first outflow launches at $\simeq4\kyr$ after the onset of gravitational collapse. The initial launch event is triggered when the increasing magnetic field strength at the pole equals the level where the local Alfv\'en velocity reaches the infall velocity, making the gas flow sub-Alv\'enic. Previously, the infalling gas was traveling much faster than the magneto-acoustic signal velocity and in turn the magnetic field topology was determined by the gas movement. When the flow approaches the sub-Alfv\'enic regime, the magnetic field lines start to resist the gas movement to a larger degree. Eventually, the Alfv\'en velocity equals the flow velocity, coinciding with a force-equilibrium between inertial and magnetic forces and the first outflow is launched by a magneto-centrifugal launching process.

\begin{figure*}
\centering
\begin{subfigure}[b]{0.36\textwidth}
        \includegraphics[width=\textwidth]{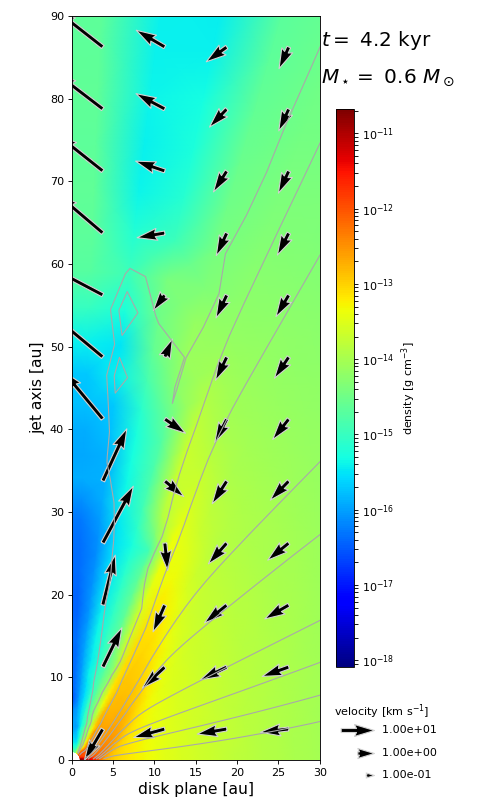}
        \caption{} 
\end{subfigure} \quad
\begin{subfigure}[b]{0.36\textwidth}
        \includegraphics[width=\textwidth]{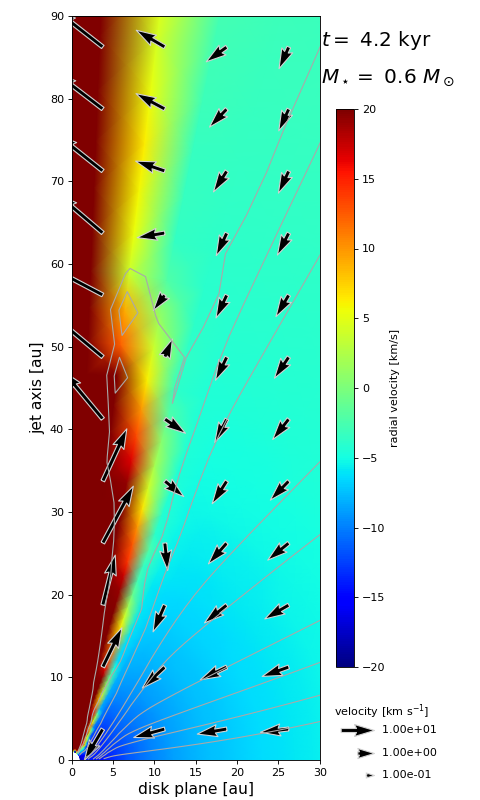}
        \caption{} 
\end{subfigure} \quad
\begin{subfigure}[b]{0.36\textwidth}
        \includegraphics[width=\textwidth]{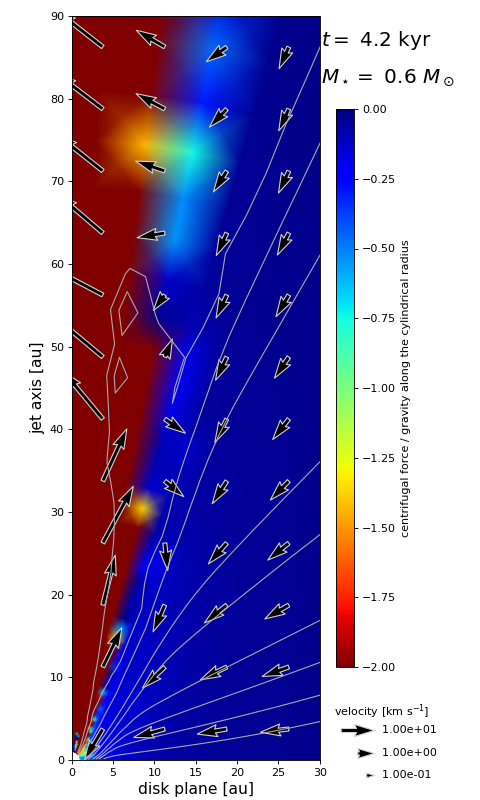}
        \caption{} 
\end{subfigure} \quad
\begin{subfigure}[b]{0.36\textwidth}
        \includegraphics[width=\textwidth]{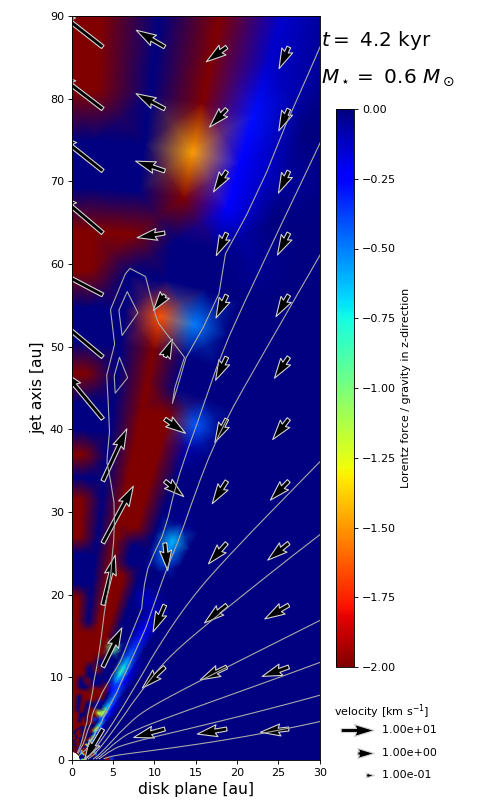}
        \caption{} 
\end{subfigure}
\caption{Jet launching from the inner accretion disk with a display of the forces involved.
	(a) Density structure.
	(b) Radial velocity.
	(c) Centrifugal force divided by gravity.
	(d) Upward pointing Lorentz force divided by gravity.
}
\label{fig:forces_during_launch}
\end{figure*}

The outflow immediately carves out an outflow cavity with a density contrast of $10^{2}$. This also changes the magnetic field topology in the launching region, and the topology of the outflow region changes rapidly for $\simeq 100\yr$, until it reaches a state where it does not change significantly on timescales on the order of several hundred years. 
The magnetic field lines that provide the necessary rotational acceleration are anchored in the inner accretion disk at radii $<1\au$.\footnote{We simulate the rapid rotation of the protostar and the inner accretion disk by zero-gradient boundary conditions for the rotational component of the velocity of the sink cell. 
}
In our simulations, the acceleration happens far above the surface of the disk and with magnetic field lines steeply inclined with respect to the equatorial plane, as visible in Fig. \ref{fig:forces_during_launch}.
This is not contradicting \citet{Blandford1982}, who state that in their solution for magneto-centrifugally launched jets the angle between magnetic field lines and the surface of the accretion disk has to be $< 60\degree$ to be able to accelerate a particle along the corresponding field line.
They assumed perfect conductivity, self-similarity for the properties of the jet, and that the particle is launched from the surface of a thin disk. All these assumptions are not met in our simulations.
However, this is unsurprising as \citet{Blandford1982} already argued that ``a realistic accretion disk must be far more complicated.''
In our evolving collapse, the foot points of magnetic field lines can move with the accreting material inwards at the equatorial plane, while particles higher up on these field lines are potentially still infalling. When this process continues, the rotational velocity at the foot point of the field line increases, and the rotating field line then accelerates the rotational movement of the previously infalling material, eventually increasing the centrifugal force to match gravity and ultimately reverse the flow direction along the field line.
Another process that contributes material to the launching region above the disk and works at the same time is nonideal MHD, which enables particles to slip with respect to the magnetic field lines. When the infalling gas of the envelope comes to a stop at the dense region close to the disk and the protostar, due to the increasing magnetic and centrifugal forces counteracting gravity, it cannot reverse its flow direction immediately, since the high ram pressure of the infalling gas at higher altitudes prevents that. However, it can slip with respect to the field lines and can, therefore, diffuse through the wall of the outflow cavity and be accelerated upwards.
\begin{figure}
\centering
        \includegraphics[width=0.49\textwidth]{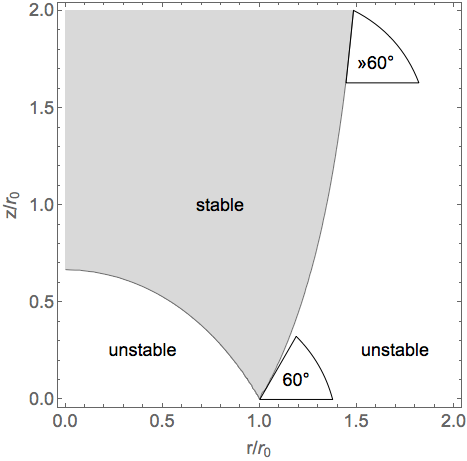}
\caption{Reproduction of Fig. 1 from \citet{Blandford1982}: equipotential surfaces for a bead on a wire, corotating with the Keplerian angular velocity $\sqrt{GM/r_0^3}$ which is released at $r_0$.}
\label{fig:Blandford_equipot}
\end{figure}
This launching at higher altitudes is even implicitly included in the model of \citet{Blandford1982}. With the magneto-centrifugal launching mechanism shown there, it is possible to accelerate particles at higher angles of $\vec B$ with the equatorial plane, when the assumption that the particle is directly launched from a thin accretion disk is dropped. This principle is visible in Fig. \ref{fig:Blandford_equipot}, showing the equipotential surfaces of a bead on a wire, where the potential $V(r,z)$ in the vicinity of a central object of mass $M$ is given by
\begin{equation}
        V(r,z) = - \frac{GM}{r_0} \left( \frac{1}{2} \left(\frac{r}{r_0}\right)^2 + \frac{r_0}{(r^2+z^2)^{1/2}} \right)\,.
\end{equation}
Here $r$ denotes the cylindrical radius and $z$ the altitude above the equatorial plane.
Following the field line anchored at $r_0$ upwards to higher $z$, we see that the angle increases continuously. This is where, in our simulation, particles are accelerated at altitudes several times the foot point radius of the field line, and with high angles of $B$.
This fast jet is stable during the course of the subsequent simulation.
Though, since the mass of the protostar and its immediate surroundings increase with time, the gravitational force increases more rapidly than the Lorentz force needed to accelerate particles upwards along the magnetic field lines. This means that the region where the Lorentz force in the vertical direction is stronger than gravity moves upwards during the course of the simulation, effectively moving the launching region with it.
There, much of the material that feeds the jet stems directly from the infalling envelope. It is channeled into the jet without ever reaching the midplane and therefore it still has significant momentum towards the protostar before its vertical velocity can be inverted by magneto-centrifugal acceleration. In turn, the region of positive vertical velocity is several astronomical units above the region of force equilibrium between magnetic and gravitational force. While, initially, it lies within $1\au$ altitude, at $15\kyr$, it has already moved to $1.8\au$ and continues to increase in altitude to $12.9\au$ at $46\kyr$.

Forty seven kiloyears after the initial collapse, the supply of material falling in from cylindrical radii close to the sink cell and with it the supply of angular momentum have reduced so much that this fast magneto-centrifugal launching mechanism ceases to work in the highest resolved simulation $126\times20$, as the Alfv\'en limiter produces too much mass, while conserving (angular) momentum. Therefore, the last remnants of the outflow then rapidly move outwards along the polar axis and in its wake, material starts to fall in again that is never accelerated upwards by magneto-centrifugal launching. As mentioned above, the simulation with lower resolution and $64\times10$ cells does not show this behavior, as the Alfv\'en limiter produces much less mass.
In all simulations, there is another outflow component that remains, a slower wide-angle wind powered by a magnetic tower, which is discussed in the following section.

\subsection{Wide-angle winds}
\label{sec:wide_angle_winds}

\begin{figure*}
\centering
\begin{subfigure}[b]{1.0\textwidth}
        \includegraphics[width=0.32\textwidth]{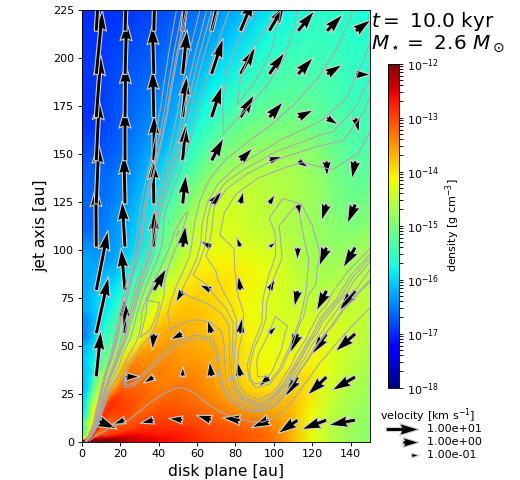}
        \includegraphics[width=0.32\textwidth]{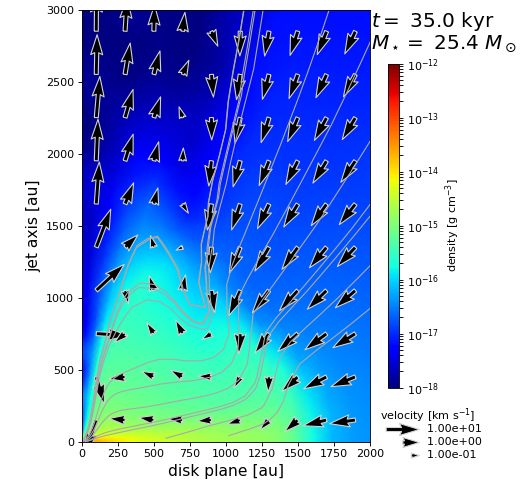}
        \includegraphics[width=0.32\textwidth]{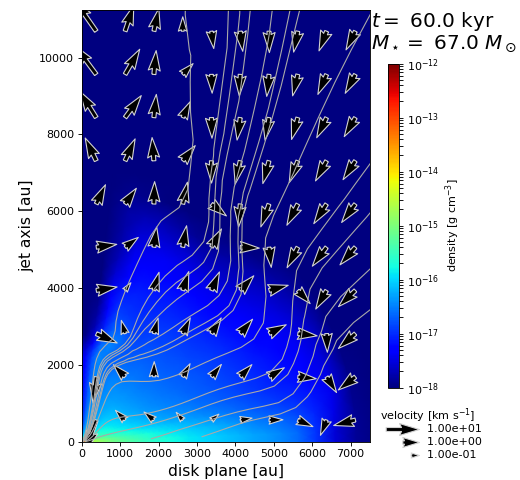}
        \caption{} 
\end{subfigure}
\begin{subfigure}[b]{1.0\textwidth}
        \includegraphics[width=0.32\textwidth]{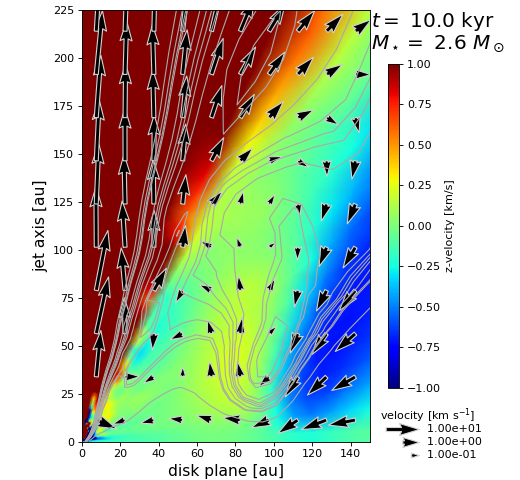}
        \includegraphics[width=0.32\textwidth]{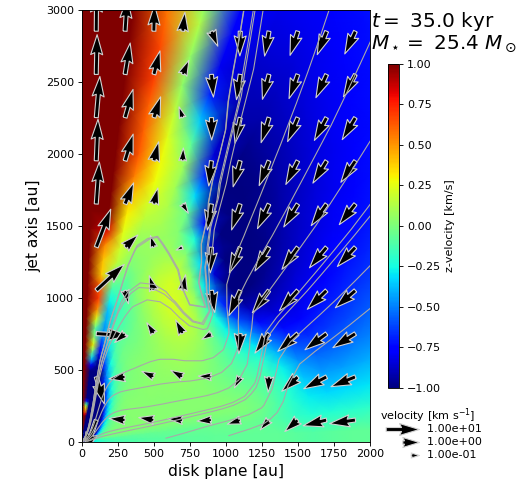}
        \includegraphics[width=0.32\textwidth]{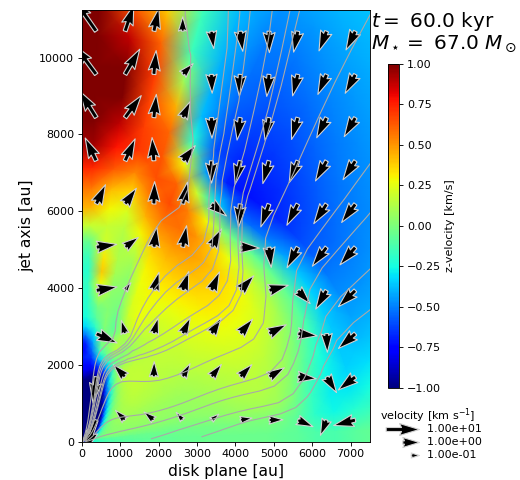}
        \caption{} 
\end{subfigure}
\begin{subfigure}[b]{1.0\textwidth}
        \includegraphics[width=0.32\textwidth]{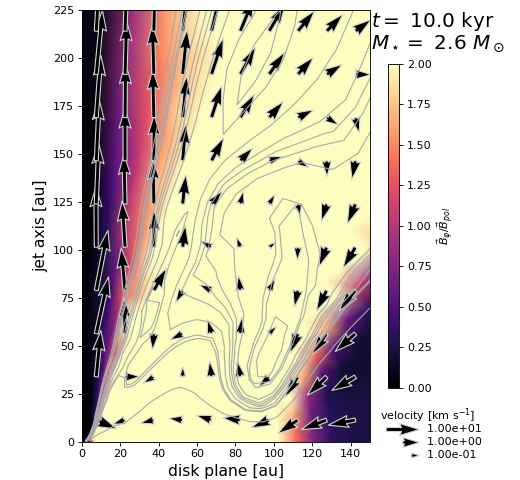}
        \includegraphics[width=0.32\textwidth]{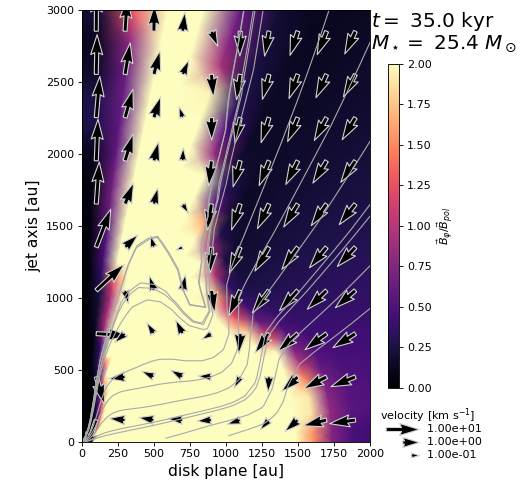}
        \includegraphics[width=0.32\textwidth]{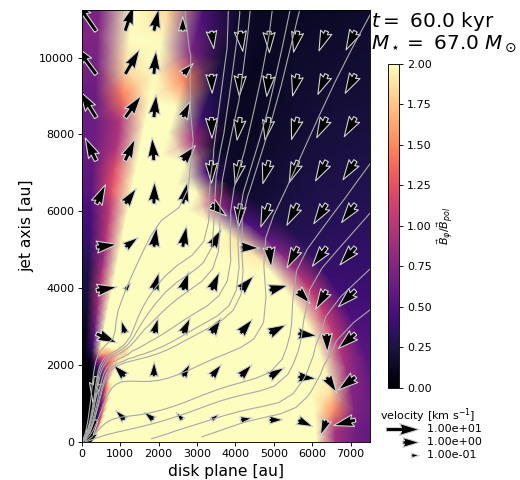}
        \caption{} 
\end{subfigure}
\caption{Evolution of the outflow, from $10\kyr$ on  the left, $35\kyr$ in the center, to $60\kyr$ to the right. 
	(a) Density structure.
	(b) Vertical velocity.
	(c) $B_\phi/B_\mathrm{pol}$.
}
\label{fig:wide_angle_winds}
\end{figure*}

The first traces of a slower wide-angle wind become apparent after $\simeq 7\kyr$. It can be ditinguished from the magneto-centrifugal jet by a strong toroidal magnetic field, $\simeq 0$ velocities in the equatorial plane, and higher densities in the launching regions. Its launching mechanism requires that magnetic field lines wind up and produce a vertical magnetic field gradient.
This can only happen if the inertial forces of the gas flow are stronger than the magnetic forces and in a region of differential rotation or with a vertical rotational velocity gradient.
Therefore, this launching mechanism develops just outside of the Alfv\'en surface and above the accretion disk, where these conditions are met for the first time. Initially, this is close enough to the magneto-centrifugal jet that both outflows merge at larger altitudes, meaning that both launching/acceleration principles contribute to the flux. However, with the growing accretion disk, the potential launching area of this magnetic tower grows towards larger radii.

Its time evolution is visible in Fig. \ref{fig:wide_angle_winds}, which shows multiple quantities at different simulation times. The first column shows the situation after the jet and tower flow have just formed but are already matured to a stable state and are clearly distinguishable at $10\kyr$. The second column shows the situation at an intermediate time, where the launching region of the tower flow has grown outwards at $35\kyr$ (please note the scale on the $x$-axis), and the last column shows the tower flow after one free-fall time, when the infall has subsided and the tower flow has expanded at $60\kyr$.

At $10\kyr$, the tower flow is clearly visible as a separate structure in the (slow) vertical velocity in panel (b), as well as through the ratio of toroidal to poloidal magnetic field\footnote{\textit{Poloidal} means the components of the magnetic field that lie in the planes perpendicular to the equatorial plane. In cylindrical coordinates, one would speak of the radial direction (often depicted by $\varpi$) and the vertical direction (usually $z$). In contrast, the toroidal component of the magnetic field corresponds to the components in the equatorial plane, in cylindrical or spherical coordinates, the azimuthal direction (usually $\phi$).} $B_\phi/B_\mathrm{pol}$ shown in panel (c). 
Here, a strong toroidal magnetic field component, shown by the bright areas in panel (c) indicates that the magnetic field lies to a large degree parallel to the equatorial plane. Usually, this happens outside of the Alfv\'en surface when the magnetic field lines are wound up by the differential gas flow of an accretion disk and hints at the magnetic pressure-driven nature of the outflow.

During the course of the subsequent simulation, the launching radius of the tower flow moves continuously outwards, as visible in Fig. \ref{fig:wide_angle_winds} when comparing the earlier images on the left to the later ones on the right.
In panel (a), we can see that the tower flow is active in areas of substantial density. Consequently, it can also accelerate a lot of mass and momentum, albeit, as visible in panel (b), to much lower terminal velocities compared to the magneto-centrifugal outflow. 
When the simulation reaches one free-fall-time, there is no more material left to fall in from above. This also means that the considerable ram-pressure of the infall is removed. Thereby, even a small vertical magnetic pressure gradient suffices to drive an outflow. As a result, the wide-angle component quickly expands up to more than $10000\au$.
Comparing the snapshot at $35\kyr$ with the one at $60\kyr$, we can see the effect of the vanishing ram-pressure of the infalling envelope, not only in the extent of the outflow but also in the higher velocities visible in panel (b) where the tower flow has finally accelerated material to velocities of several kilometres per second, while before it was only reaching fractions of a kilometre per second.

Comparing our findings with earlier simulations, for example the well-cited work by \citet{Banerjee2007}, we are, to our knowledge, the first that can clearly distinguish between magneto-centrifugal jet and the massive, slow tower flow which they also describe. In principle, we think that other simulations, for example \citet{Matsushita2017}, should be able to resolve this as well, though they do not mention a distinction in particular. 

\citet{Seifried2012} analyzed the jet-launching mechanism in detail. Similar to our results, they find outflow components that can be described as magneto-centrifugally launched and magnetic pressure-driven. 
They find a magnetic pressure-driven component in two distinct situations. In the early stages of disk formation, a magnetic tower develops but only as an early transient phenomenon. Later, magnetic pressure contributes to the launching at larger radii, and correctly they argue that in their simulations this magnetic pressure driving should be considered a part of the magneto-centrifugal launching process, which also matches the \citep{Blandford1982} model.
We agree that at larger radii, outside the Alfv\'en radius, magnetic pressure potentially contributes to the acceleration of formerly magneto-centrifugal jets. However, we find that one needs a significantly higher resolution to distinguish between magneto-centrifugal jet and tower flow. In fact, we find a similar hybrid magneto-centrifugal and magnetic pressure-driven outflow for our low-resolution simulations.

\begin{figure}
\centering
\includegraphics[width=0.5\textwidth]{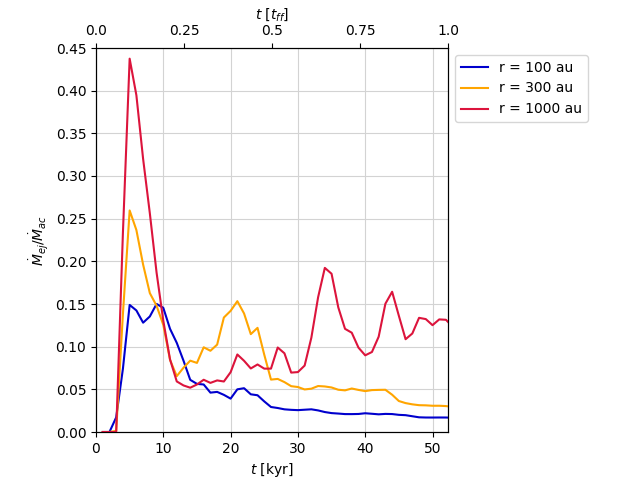}\\
\caption{The ejection-to-outflow efficiency for different radii.}
\label{fig:ejection-to-accretion-efficiency}
\end{figure}

Both launching mechanisms contribute to the ejection-to-outflow efficiency. In Fig. \ref{fig:ejection-to-accretion-efficiency}, we present the ejection-to-outflow efficiency at different radii, for simulation $64\times10$. At all radii, the launching event with its very massive outflow by entrainment is visible. However, since more material can be entrained at higher radii, the ejected mass gets larger with increasing radius. After this initial ejection by entrainment, the efficiency goes down to values $\simeq0.1$ as is expected from observations.
Towards even later evolutionary timescales, we can observe how the magneto-centrifugal jet and the tower flow separate from each other in radius. At one free-fall time, the magneto-centrifugal jet component that is active well within the first $100\au$ only reaches a low ejection-to-accretion ratio of $0.015$, while at larger radii the tower flow contributes significantly, increasing the ejection-to-outflow efficiency to $0.15$ at $1000\au$.

While the magneto-centrifugal jet is affected by the Alfv\'en limiter in the highest-resolution simulations (which we do not discuss here), we want to note that integrated over the whole timescale of the simulation, we find ejection-to-outflow efficiencies of $\simeq0.12$, which are quite compatible with observations. This is visible in panels (a)
and (b) of Fig. \ref{fig:conv_star}. The reason is that the magneto-centrifugal jet only contributes to a minor degree to the total outflowing mass, additionally indicating that the total observed ejection-to-outflow efficiency is also a time-dependent quantity.

\section{Caveats}
\label{caveats}
The adopted mass-to-flux ratio of $\mu=20$ is relatively high in comparison to the observational upper limit, suggesting $\mu\leq2$. Correspondingly, our magnetic field strength is a factor of $\simeq 10$ lower than many observations suggest. 
We chose this lower mass-to-flux ratio because we expect that the redistribution by nonideal effects is crucial for the realistic evolution of the system. While we do include ohmic dissipation to simulate the magnetically decoupled dead zone in the accretion disk, we do not consider nonideal effects that work on larger scales and at lower densities, such as for example ambipolar diffusion, which could reduce the total magnetic flux significantly. 
On another note, we expect the high mass-to-flux ratio to have only a minor influence on the outflows, as Ohmic resistivity will quickly redistribute the magnetic field in the inner accretion disk. Additionally, we consider a realistic magnetic field and cloud core evolution to only  be possible when including radiative transfer. Radiative transfer would enable us to estimate the ionization degree much more realistically, as well as providing a more realistic thermal pressure support in the disk, which in turn influences the magnetic field morphology. We still deem the effects of different magnetic field an essential aspect of collapse simulations and are planning to study its effects in future projects after including radiation transfer.

\section{Summary}
\label{sec:summary}

There are numerous unresolved questions regarding the formation process of massive stars. In contrast to low-mass stars, they are commonly found at large distances from our solar system, reducing the potential resolution of their observations. This is not the only factor that makes their formation process more difficult to observe than their low-mass counterparts. More massive stars form in equally massive and therefore denser cloud cores, rendering their environment more opaque to observations. There are, however, traces of their formation process that are visible to us; their bright large-scale jets and outflows, which are the focus of the present work.
Given that we have access to much more information on low-mass protostars, it is natural to pose the question of to what degree the formation processes of low and high-mass stars work similarly, and where they diverge. Another important issue arises from the fact that massive stars influence their environment via different feedback mechanisms much more than low-mass stars do, even in their early stages of evolution. Therefore, important aspects of our analyses are the properties of protostellar jets and outflows.

To contribute to the solution of some of those issues, we conduct  MHD simulations using the state-of-the-art code PLUTO, combining nonideal MHD, self-gravity, and very high resolutions, as they have never been achieved before. Our setup includes a $100\Msol$ cloud core that collapses under its own self-gravity to self-consistently form a dense disk structure, launching tightly collimated magneto-centrifugal jets and wide angle tower flows.
To benchmark the quality of our results and analyze requirements of general MHD simulations, we conduct a detailed convergence study, not only changing the resolution, but also the radius of sink cells, which represent the forming protostar, and analyze the requirements to resolve physical processes in the disk and during outflow launching.

Previous simulations sometimes found that centrifugally supported disk formation is even possible in ideal MHD. We do not believe this to be accurate.
We conclude that nonideal MHD is required to form such accretion disks. We think that in earlier studies, which show the formation of small circumstellar disks in ideal MHD, the formation process is most likely enabled by numerical diffusivity due to lack of resolution per disk pressure scale height. With the resolution our simulations provide, centrifugally supported disk formation is only possible when considering a magnetically dead zone in the most dense regions of the accretion disks. By including such a dead zone, the size of the disk in our MHD simulations approaches values of nonmagnetic simulations, that is, several $100\au$, where the exact value depends on the initial angular momentum content of the collapsing cloud.
The resolution study reveals a strong dependence of the disk size on spatial resolution, with higher-resolution simulations forming larger and more massive disks. The reason is that it is necessary to resolve the densest part of the disk's midplane in order to see the strongest impact of the nonideal MHD term. We found that a converged result is obtained for a spatial resolution of $\Delta x \leq 0.17 \au$  (at a radius of $1\au$).

The central result of the resolution study is that the nature of the outflow depends critically on spatial resolution. Only high-resolution simulations are able to resolve a magneto-centrifugally launched, and highly collimated jet from a slow wide-angle magnetic pressure-driven tower flow. As reported in earlier simulations, we found a broad and massive tower flow at early stages of the cloud collapse, though only when using low resolutions.
In higher-resolution simulations, this massive outflow separates into two distinguishable, temporally and spatially separated components. The magneto-centrifugal jet is active close to the polar axis at early times. Later, the tower flow develops at larger disk radii.
In agreement with earlier simulations, we found that the majority of the angular momentum is transported by the tower flow, since it is active at larger radii where more angular momentum is present in the accretion disk. Therefore, we conclude that another mechanism must be available to remove angular momentum before the tower flow has reached its full potential. A very good candidate is gravitational torques, as described by \citet{Kuiper2011}.

The convergence study, which varies the size of the sink cells, representing the accreting protostar, shows converged results for sink-cell sizes $\leq 3.1\au$. For large sink cells, the magnetic field morphology in the jet launching regions deviates significantly from the morphology evolved in simulations with sink-cell sizes of $\leq 3.1\au$, leading to a much broader magneto-centrifugal outflow that also, due to entrainment, transports much more mass. The resulting wide, and massive outflow partially hinders accretion, and, as a result, these simulations show temporally decreased accretion rates to the protostar.
On the other hand, two simulations, with sink-cell sizes of $1.0\au$ and $3.1\au,$ respectively, show a close to identical evolution of the magnetic field morphology, resulting in a converged disk, jet, and tower-flow system with higher and potentially more realistic accretion rates.

Our force analyses show that the launching, acceleration, and collimation of the jet component is consistent with the mechanism described by the analytical work of \citet{Blandford1982}. Similarly, the magnetic tower flow follows the principles of \citet{Lynden-Bell2003}.
In addition to these analytical and idealized studies, our cloud-collapse simulations include the interaction of the two outflow components with the large-scale infalling stellar environment and provide a temporal dimension that shows a conclusive picture of a cloud-collapse with great dynamic range.

By resolving these high dynamic ranges and the two separate outflow components, we found that the mass outflow rate is dominated by the entrained material from the interaction of the jet with the stellar environment and that only parts of the ejected medium have been directly launched from the accretion disk. Furthermore, accretion as well as ejection are processes that vary in time.
Taking into account both the mass that is launched from the  surface of the disk and the entrained material from the envelope, we found an ejection-to-accretion efficiency of $10\%$ over the whole course of our simulation, as expected by observations.

Jets and outflows in star formation represent a universal process with respect to the basic physics of launching, acceleration, and collimation, and act independently of stellar mass. Massive stars not only possess slow wide-angle tower flows, transporting large amounts of mass, but also produce magneto-centrifugal jets, just as their low-mass counterparts do.
As a consequence, the very early evolution of the system acts as a scaled-up version of low-mass star formation, at least with respect to the launching of jets and outflows. The actual significant contrast between low-mass and high-mass star formation lies in the ``embeddedness'' of the high-mass star, which implies that the jet and tower flow interact with the infalling large-scale stellar environment. This potentially results in the entrainment of much more mass, which additionally leads to visually broader outflows on large scales. According to our results, the original launching mechanisms of these entrained massive outflows could still be magneto-centrifugal in nature. On the contrary, a magnetic-pressure-driven outflow would even be hindered by the massive envelope of massive protostars and can only develop fully when much of the original envelope has already dispersed.

The massive young stellar object in the LMC observed by \citet{McLeod2018} appears to possess such a highly collimated fast magneto-centrifugal jet as the one describe here. The reported jet velocities of $300-400$ km/s and derived outflow lifetimes of $(28-37)\kyr$ match our simulations very well. When only considering the magneto-centrifugal jet, that is, regions with vertical velocities of $>10$ km/s, we arrive at an average mass outflow rate of $2.15\times 10^{-6}\Msol$ yr$^{-1}$, which is remarkably close to their estimate of $2.9\times 10^{-6}\Msol$ yr$^{-1}$.

\begin{acknowledgements}
We want to express our gratitude to Christian Fendt, Ralf Pudritz, Neal Turner, Harold Yorke, Andrea Mignone, Shu-Ichiro Inutsuka, and Daniel Seifried, for the fruitful discussions and their helpful thoughts on our project. Also, we want to thank the referee for his/her constructive comments.
We acknowledge financial support via the Emmy Noether Research Programme funded by the German Research Foundation (DFG) under grant no. KU2849/3-1.
Our simulations were performed on the computational resources ForHLR I and ForHLR II funded by the Ministry of Science, Research and the Arts Baden-W\"urttemberg and DFG.
\end{acknowledgements}


\bibliographystyle{aa}
\bibliography{literature}

\end{document}